\documentclass[showpacs, superscriptaddress,amsmath,amssymb, aps]{revtex4-1}

\usepackage[T1]{fontenc}
\usepackage[latin9]{inputenc}
\usepackage{verbatim}
\usepackage[latin9]{inputenc}
\usepackage{verbatim}
\usepackage{amsmath}
\usepackage{amssymb}
\usepackage{graphicx}
\usepackage{wrapfig}
\usepackage{esint}
\usepackage{dcolumn}
\usepackage{bm}
\usepackage{setspace}
\usepackage{color}
\usepackage{soul}
\usepackage[T1]{fontenc}
\usepackage{lmodern}
\usepackage{subfig}
\usepackage{amsmath}
\usepackage{amssymb}
\usepackage{graphicx}
\usepackage{esint}

\begin{document}
\preprint{APS/123-QED}

\title{Ruderman-Kittel-Kasuya-Yosida (RKKY) interaction in Weyl semimetals with tilted energy dispersion}
\author{Anirban Kundu}
\affiliation{Department of Electrical and Computer Engineering, National University of Singapore, 4 Engineering Drive 3, Singapore 117576}
\affiliation{Physics Department, Ariel University, Ariel 40700, Israel}
\author{M. B. A. Jalil}
\affiliation{Department of Electrical and Computer Engineering, National University of Singapore, 4 Engineering Drive 3, Singapore 117576}
\date{\today}


\begin{abstract}
Ruderman-Kittel-Kasuya-Yosida (RKKY) is an essential long range magnetic
interaction between magnetic impurities or magnetic layered structures,
the magnitude of which oscillates with the distance ($\mathrm{R}$)
between them. We have investigated the RKKY interaction between two
magnetic impurities in both time-reversal and inversion symmetry broken
Weyl semimetals (WSMs) where the energy dispersion is tilted in momentum
space and the momentum of the conduction electron is locked with the
pseudo-spin. Two important features are revealed, firstly, at the
small tilt limit, we show that the RKKY coupling varies quadratically
with the tilt parameter and strikingly, at large separation distance
$\mathrm{R}$, the coupling decays as $1/\mathrm{R}$ compared to
the conventional of $1/\mathrm{R}^{3}$ dependence exhibited by WSMs
with non-tilted dispersion. The slower decay by two orders i.e. ($1/\mathrm{R}$
as opposed to $1/\mathrm{R}^{3}$) of the RKKY coupling is significant
for maintaining long range RKKY coupling. Secondly, the RKKY coupling
exhibits an anisotropy with respect to the angle between the tilt
direction ($\mathbf{w}$) and the separation direction $\mathbf{R}$
unlike the case of non-tilted WSMs which exhibit isotropic RKKY coupling.
Consequently, the RKKY coupling in tilted WSMs alternately favours
ferromagnetic and anti-ferromagnetic orders and vice-versa with the
change of the angle. Our results are derived analytically and verified
by numerical calculations based on realistic parameter values. 
\end{abstract}

\maketitle

%
%

\section{Introduction}

Recent progress on the study of various exotic topological materials
and their transport properties has made Weyl semimetals a fascinating
subject to study. A Weyl semimetal (WSM) can be produced from a four-fold
degenerate Dirac semi-metal by breaking either or both of the time-reversal
(TRS)/inversion-broken (IR) symmetry \cite{1367-2630-9-9-356,PhysRevB.83.205101}.
In Dirac semi-metals, the band touching points known as nodes are
four-fold degenerate and protected by TRS and IR symmetries. By breaking
at least one of these symmetries, a pair of Weyl nodes is produced.
The nodes harbor opposite topologically protected chiral charge $\eta$
($\eta=\pm1,\pm2,...$) which can be determined by calculating the
Berry curvature $\boldsymbol{\Omega}_{\mathbf{k}}$ ($\varointclockwise\text{d}\boldsymbol{s}\cdot\boldsymbol{\Omega}_{\mathbf{k}}=2\pi\eta$,
$\boldsymbol{\Omega}_{\mathbf{k}}=i\nabla\times\left\langle u_{\mathbf{k}}|\nabla_{\mathbf{k}}u_{\mathbf{k}}\right\rangle $,
where $\left|u_{\mathbf{k}}\right\rangle $ is the Bloch wave function)
\cite{PhysRevLett.49.405,PhysRevB.83.205101}. The electronic dispersion
and transport phenomena associated with WSMs have been explored by
a number of experimental \cite{Arnold2016,Li2015,Li2016,PhysRevX.5.031023,Xiong413,Zhang2016}
and theoretical studies \cite{PhysRevB.96.115202,PhysRevB.96.085114,PhysRevB.99.115121,PhysRevLett.119.176804,Kundu_2020,PhysRevApplied.14.034007,Yesilyurt2016}.
The distinct topology of the WSM leads to various unique electronic
and transport properties such as, Fermi arc surface states connecting
a pair of Weyl nodes, Adler--Bell--Jackiw (ABJ) anomaly and negative
magentoresistance \cite{PhysRevLett.124.116802,PhysRevB.100.085424,PhysRevB.92.121109,PhysRevB.92.041107,PhysRevB.86.195102}.
Besides, WSMs also exhibit superconducting phases \cite{PhysRevLett.113.046401,Qi2016,Bachmanne1602983,PhysRevLett.114.096804},
and the chiral Hall effect \cite{PhysRevLett.113.187202,Wang2018}.
The WSM phase was first observed in the candidate material TaAs \cite{PhysRevX.5.031013,Lv2015,Xu613}
and later NbAs, TaP and NbP \cite{Xu2016,Huang2015}. 

An important aspect of the study of WSM materials is their magnetic
ordering. WSMs lack spontaneous magnetic ordering originating from
the Heisenberg interaction. However magnetic interactions may be generated
between magnetic impurities through an indirect exchange coupling
mechanism namely the Ruderman-Kittel-Kasuya-Yoshida (RKKY) interaction.
For a system with magnetic impurities, the coupling is generated from
the Heisenberg exchange interaction between the impurities and the
conduction electrons near the Fermi level, which can be modeled by
the second-order perturbation theory. 

Several theoretical works have been made to address the long-range
RKKY type magnetic interaction in WSMs. A study of the RKKY interaction
for both the TRS breaking and IR breaking WSMs has been done by Ref.
\cite{PhysRevB.92.241103}, but, in the absence of any dispersion
tilt. This study reveals the presence of only the conventional long-range
Heisenberg type coupling. In the presence of tilt and for the TRS-breaking
WSMs the RKKY interaction between the magnetic impurities in WSM has
been studied in Ref. \cite{PhysRevB.99.165111}. The work assumes
the presence of real-spin-momentum locking and considers only the
special case where the tilt direction is fixed along the vector connecting
the two Weyl nodes. However, most of the WSMs experimentally studied
so far are found to be IR-breaking. Besides, these WSM systems lack
real spin-momentum locking and instead exhibit pseudo-spin momentum
locking, as described by most of the minimal energy Hamiltonians of
WSM in the literature \cite{PhysRevX.5.031013,Lv2015,Xu2016}. 

In the conventional RKKY interactions, e.g., in diluted magnetic semiconductors
\cite{PhysRevB.68.235208,PhysRevB.70.075205,doi:10.1063/1.5097673},
in topological insulators \cite{Ho_2020,doi:10.1063/1.4977072}, the
RKKY interaction is spatially isotropic and is a function only of
the separation distance between the magnetic impurities. By contrast,
in this paper we will consider the RKKY interactions in both the TRS
and IR breaking WSMs where the isotropy is broken by the presence
of tilt in the energy dispersion. Thus, in our system there are three
directions under consideration: the direction of the tilt, the separation
vector between two impurities and the vector joining the two Weyl
nodes. We will explore the directional dependence of the RKKY interaction
as a result of the interplay of these three special directions, and
utilize the corresponding degrees of directional freedom to modulate
the RKKY coupling and switch its sign i.e., from ferromagnetic to
anti-ferromagnetic coupling and vicce-versa.

In this manuscript, we will investigate the effect of dispersion tilt
on the RKKY coupling for two special cases: (1) when time-reversal
symmetry (TRS) is broken but inversion symmetry (IR) is intact, i.e.,
when a pair of WPs are separated in momentum space ($\mathbf{Q}$)
but the nodes appear at the same energy and (2) when IR is broken
but TRS is intact i.e. when WPs are separated along energy axis ($\epsilon_{0}$).
In addition, we will consider the contributions to the RKKY interaction
which originate from the scattering between any two energy states
within a particular band, namely intra-band scattering, as well as
scattering between different bands with either the same or opposite
chiralities, namely inter-band scattering.

\section{Method}

The Hamiltonian for a tilted WSM around a pair of Weyl nodes is \cite{Zyuzin2016,PhysRevLett.117.077202,PhysRevLett.117.086401,PhysRevB.92.241103,PhysRevB.92.224435},

\begin{equation}
H_{0}=\eta\hbar v_{F}\boldsymbol{\tau}\cdot\left(\mathbf{k}-\eta\mathbf{Q}\right)+\hbar\mathbf{w}_{\eta}\cdot\left(\mathbf{k}-\eta\mathbf{Q}\right)\tau_{0}+\eta\tau_{0}\epsilon_{0},\label{eq: unperturbed Hamiltonain}
\end{equation}
where, $\eta=\pm1$ indicates the chirality of the Weyl nodes in the
two valleys, $2\mathbf{Q}$ is the separation between them in the
momentum space and $2\epsilon_{0}$ is the separation along the energy
axis, $\boldsymbol{\tau}$ is the Pauli vector acting on the chirality
space and $\mathbf{w}_{\eta}$ is the tilt vector for the Weyl nodes
with chirality $\eta$. The tilt serves as an additional anisotropy
term in the Hamiltonian \cite{PhysRevB.78.045415}. The corresponding
energy dispersion is as follows, $\epsilon_{\mathbf{k},\eta}=\eta v_{F}\left|\mathbf{k}-\eta\mathbf{Q}\right|+\mathbf{w}_{\eta}\cdot\left(\mathbf{k}-\eta\mathbf{Q}\right)+\eta\epsilon_{0}$.

To calculate the RKKY interaction, we consider two impurities with
magnetic moments $\mathbf{S}_{1}$ and $\mathbf{S}_{2}$ and situated
at $\mathbf{R}_{1}$ and $\mathbf{R}_{2}$. The s-d interaction between
the conduction electrons and the two impurities can be regarded as
a perturbation term, $H'(\mathbf{r})=J\text{ }\sum_{i=1,2}\boldsymbol{S}_{i}\cdot\boldsymbol{\sigma}\delta(\mathbf{r}-\mathbf{R}_{i})$,
where $\mathbf{r}$ represents position coordinate and the $\boldsymbol{\sigma}$ corresponds to Pauli vector representing
real spin degrees of freedom. The RKKY interaction can be obtained
by using the second-order perturbation theory \cite{PhysRevB.69.121303}.
The second order energy correction due to the above perturbation is,
\begin{equation}
\epsilon^{(2)}=\sum_{\mathbf{k}'\eta'(\neq\mathbf{k}\eta)}\frac{|<\mathbf{k}\eta|H'(\mathbf{r})|\mathbf{k}'\eta'>|^{2}}{\epsilon_{\mathbf{k}'\eta'}-\epsilon_{\mathbf{k}\eta}}\left(f_{\mathbf{k}'\eta'}-f_{\mathbf{k}\eta}\right),
\end{equation}
where $f_{\mathbf{k}'\eta'}$ and $f_{\mathbf{k}\eta}$ are the Fermi
functions corresponding to the two valleys $\eta$ and $\eta'$. While
carrying out the summation above, we separate the contribution from
the intra-valley transition, i.e., transition between two states within
a valley with a particular chirality ($\eta=\eta'$) and the inter-valley
transition between two valleys with opposite chiralities ($\eta\neq\eta'$).
With this we rewrite the above term as follows, 
\begin{equation}
\epsilon^{(2)}=2\left(\boldsymbol{S}_{1}\cdot\boldsymbol{S}_{2}\right)F\left(\mathrm{R}\right),
\end{equation}
where, $\mathrm{R}=|\mathbf{R}|=|\mathbf{R}_{1}-\mathbf{R}_{2}|$
and $F\left(\mathrm{R}\right)$ is typically known to be the RKKY
range function which is given by,

\begin{equation}
F\left(\mathrm{R}\right)=\sum_{\eta}F_{\text{intra}}^{\eta}\left(\mathrm{R}\right)+\sum_{\eta}e^{i\left(\eta2\mathbf{Q}\cdot\mathbf{R}\right)}F_{\text{inter}}^{\eta,-\eta}\left(\mathrm{R}\right).\label{eq: total range function}
\end{equation}
In the above, we separate out the intra and inter-valley contributions.
The intra-valley contribution in the above can be given as (after
changing $\sum_{\mathbf{k}'\eta'(\neq\mathbf{k}\eta)}\rightarrow\int\text{d}\mathbf{k}\text{ }\int\text{d}\mathbf{k}'$),
\begin{align}
F_{\text{intra}}^{\eta}\left(\mathrm{R}\right) & =\left(\frac{1}{\hbar v_{F}}\right)\text{Re}\int_{{k}=0}^{k_{F\eta}}\text{d}\mathbf{k}\text{ }\int_{{k}'=k_{F\eta}}^{\infty}\text{d}\mathbf{k}'\frac{e^{i\left(\mathbf{k}'-\mathbf{k}\right)\cdot\mathbf{R}}}{\left(k'-k\right)+\frac{\mathbf{w}_{\eta}}{v_{F}}\cdot\left(\mathbf{k}'-\mathbf{k}\right)},\label{eq:range function intra}
\end{align}
where $k=|\mathbf{k}|$, $k'=|\mathbf{k}'|$ and $k_{F\eta}$ is the Fermi wave vector corresponding to a valley
with chirality $\eta$ and is defined by, $\epsilon_{F}=\hbar v_{F}k_{F\eta}+\eta\epsilon_{0}$.
Similarly, the inter-valley contribution can be expressed as, 
\begin{equation}
\text{ }F_{\text{inter}}^{\eta,-\eta}\left(\mathrm{R}\right)=\left(\frac{1}{\hbar v_{F}}\right)\text{Re}\int_{{k}=0}^{k_{F\eta}}d\mathbf{k}\text{ }\int_{{k}'=k_{F,-\eta}}^{\infty}d\mathbf{k}'\frac{e^{i\left(\mathbf{k}'-\mathbf{k}\right)\cdot\mathbf{R}}}{\left(k'-k\right)+\left(\frac{\mathbf{w}_{-\eta}}{v_{F}}\cdot\mathbf{k}'-\frac{\mathbf{w}_{\eta}}{v_{F}}\cdot\mathbf{k}\right)-\eta2\epsilon_{0}}\label{eq:range function inter}
.\end{equation}
We calculate the integrations in above two functions (Eqs. (\ref{eq:range function intra})
and (\ref{eq:range function inter})) for different types of WSMs.
In usual RKKY calculation the oscillatory decaying range function
originates from the singularity (poles) in the denominator. In the
absence of tilt ($\mathbf{w}_{\eta}=0$) the pole occurs at $k'=k$
in Eq. (\ref{eq:range function intra}) and $k'=k+\eta2\epsilon_{0}$
in Eq. (\ref{eq:range function inter}). However, in the presence
of a finite tilt, the poles occur at $\mathbf{k}'=\mathbf{k}$, i.e.,
the solution for the poles depends on the directions of the wave vectors.
In this case, it is no longer possible to solve the integral analytically
when the tilt $\mathbf{w}$ and the separation vector $\mathbf{R}$
are aligned along with arbitrary directions. The analytical expression
for the integral can only be obtained by considering the small tilt
limit and expanding the expressions in the integrals of Eqs. (\ref{eq:range function intra})
and (\ref{eq:range function inter}) up to the second-order in tilt
parameter. Finally, we consider two types of tilt in our analysis,
parallel tilt i.e., when \textit{$\mathbf{w}_{\eta}=\mathbf{w}$}
for a pair of Weyl nodes and opposite tilt i.e., when $\mathbf{w}_{\eta}=\eta\mathbf{w}$.
For each type of tilt, we calculate the effects of tilt for both the
TRS breaking and IR breaking cases.

\section{Small tilt limit ($|\mathbf{w}_{\eta}|/v_{F}\ll1$)}

In the small tilt limit, we perform a Taylor expansion of the intra-valley
range function to the second-order in tilt (see Supplemental Information
for the detailed calculations). The intra-valley range function in
that case can be expressed as,

\begin{align}
F_{\text{intra}}^{\eta}\left(\mathrm{R}\right) & =\sum_{m=0}^{2}\left(\frac{\mathbf{w}_{\eta}}{v_{F}}\cdot\frac{\partial}{\partial\mathbf{R}}\right)^{m}\chi^{(m)}\left(\mathrm{R}\right),\label{eq:Fintramain}
\end{align}
where, 
\begin{equation}
\chi^{(m)}\left(\mathrm{R}\right)=\left(\frac{1}{\hbar v_{F}}\right)\text{Re}\left(-i\right)^{m}\int_{k=0}^{k_{F\eta}}\text{d}\mathbf{k}\text{ }\int_{k'=k_{F\eta}}^{\infty}\text{d}\mathbf{k}'\frac{e^{i\left(\mathbf{k}'-\mathbf{k}\right)\cdot\mathbf{R}}}{\left(k'-k\right)^{m+1}}.\label{eq:Fintramainchi}
\end{equation}
Performing a similar expansion of the inter-valley range function,
\begin{align}
F_{\text{inter}}^{\eta,-\eta}\left(\mathrm{R}\right) & =\sum_{m=0}^{2}\zeta^{(m,\eta,-\eta)}\left(\mathrm{R}\right),\label{eq:Fintermain}
\end{align}
where, 
\begin{equation}
\zeta^{(m,\eta,-\eta)}\left(\mathrm{R}\right)=\text{Re}\int_{k=0}^{k_{F\eta}}d\mathbf{k}\text{ }\int_{k'=k_{F,-\eta}}^{\infty}d\mathbf{k}'\frac{\hbar^{m}\left(\mathbf{w}_{\eta}\cdot\mathbf{k}'-\mathbf{w}_{-\eta}\cdot\mathbf{k}\right)^{m}e^{i\left(\mathbf{k}'-\mathbf{k}\right)\cdot\mathbf{R}}}{\left(\hbar v_{F}\left(k'-k\right)\mp2\epsilon_{0}\right)^{m+1}}.\label{eq:Fintermain-1}
\end{equation}
One can immediately see, the first-order term ($m=1$) in Eqs. (\ref{eq:Fintramain}
and \ref{eq:Fintermain-1}) is linearly dependent on the wave-vector
and as result would become zero after integration over the angular
part (see Supplemental materials). So, the total RKKY range function
(omitting the $1^{\text{st}}$ order in $|\mathbf{w}|$-terms) is,
\begin{align}
F\left(\mathbf{R}\right) & =\sum_{m=\{0,2\}}\sum_{\eta}\left(\frac{\mathbf{w}_{\eta}}{v_{F}}\cdot\frac{\partial}{\partial\mathbf{R}}\right)^{m}\chi^{(m)}\left(\mathrm{R}\right)+\sum_{m=\{0,2\}}\sum_{\eta}\zeta^{(m,\eta,-\eta)}\left(\mathrm{R}\right).\label{eq: FRtotal}
\end{align}
Next, we consider two types of WSMs: (a) when TRS is broken, i.e.,
the Weyl nodes are separated in $k$-space with band touching points
at the same energy ($\epsilon_{0}=0$) and (b) when IR is broken,
where the nodes are separated along energy axis ($\mathbf{Q}=0$).
For each type, we consider both cases where the tilt vectors at the
two Weyl nodes are parallel ($\mathbf{w}_{\eta}=\mathbf{w}$) and
when they are chiral or opposite to one another ($\mathbf{w}_{\eta}=\eta\mathbf{w}$).
It turns out (as will be shown later) while the RKKY range functions
are strongly dependent on the magnitude of tilt, the tilt configuration
in the two Weyl nodes (either parallel or chiral) does not play such
a significant role in determining the RKKY coupling. Hence, we will
only present the results for the parallel tilt case in the main manuscript,
while similar results for the chiral tilt case are presented in the
Supplemental section. On the other hand, the type of symmetry-breaking
would significantly affect the RKKY coupling and would be presented
separately in the main manuscript. 

\subsection{Broken TRS ($\epsilon_{0}=0$ and $\mathbf{Q}\protect\ne0$)} 

We first consider the TRS breaking case, for which the corresponding
Hamiltonian from Eq. (\ref{eq: unperturbed Hamiltonain}) is, $H_{0}=\eta v_{F}\boldsymbol{\tau}\cdot\left(\mathbf{k}-\eta\mathbf{Q}\right)+\mathbf{w_{\eta}}\cdot\left(\mathbf{k}-\eta\mathbf{Q}\right)\tau_{0}$.
Since the Fermi wave vector is the same for both Weyl nodes, the integration
limits in Eq. (\ref{eq:range function intra}) and (\ref{eq:range function inter})
simplify to, $\int_{k=0}^{k_{F}}d\mathbf{k}\text{ }\int_{k'=k_{F}}^{\infty}d\mathbf{k}'$
where $k_{F}=\epsilon_{F}/\hbar v_{F}$. For the special case of $\mathbf{w}\parallel\hat{z}$
and in the limit of small tilt, we can derive the analytical expression
of the total range function. Denoting $\zeta^{(0,\eta,-\eta)}\left(\mathrm{R}\right)=\chi^{(0,\eta)}\left(\mathrm{R}\right)$
and $\zeta^{(2,\eta,-\eta)}\left(\mathrm{R}\right)=-\left(\frac{\mathbf{w}}{v_{F}}\cdot\frac{\partial}{\partial\mathbf{R}}\right)^{2}\chi^{(2,\eta)}\left(\mathrm{R}\right)$
in Eqs. (\ref{eq: total range function}, \ref{eq:Fintramain}, \ref{eq:Fintermain}),
we obtain the following 
\begin{align}
F\left(\mathbf{R}\right) & =2\left(1+\cos\left(2\mathbf{Q}\cdot\mathbf{R}\right)\right)\left(\chi^{(0,\eta)}\left(\mathrm{R}\right)-\left(\frac{\mathbf{w}}{v_{F}}\cdot\frac{\partial}{\partial\mathbf{R}}\right)^{2}\chi^{(2,\eta)}\left(\mathrm{R}\right)\right),\label{eq: intra-total-range}
\end{align}
where,

\begin{align}
\chi^{(0,\eta)}\left(\mathrm{R}\right) & =\frac{4\pi^{3}\left(\left(3-6k_{F}^{2}\mathrm{R}^{2}\right)\cos(2k_{F}\mathrm{R})+6k_{F}\mathrm{R}\sin(2k_{F}\mathrm{R})-3\right)}{3\mathrm{R}^{5}},\\
\chi^{(2,\eta)}\left(\mathrm{\mathrm{R}}\right) & =\frac{4\pi^{3}\left(-3\left(4k_{F}^{2}\mathrm{R}^{2}+1\right)+3\left(2k_{F}^{2}\mathrm{R}^{2}+1\right)\cos(2k_{F}\mathrm{R})+6k_{F}\mathrm{R}\sin(2k_{F}\mathrm{R})\right)}{3\mathrm{R}^{3}},\\
\left(\frac{|\mathbf{w}|}{v_{F}}\frac{\partial}{\partial R_{z}}\right)^{2}\chi^{(2,\eta)}\left(\mathrm{R}\right) & =\left(\frac{|\mathbf{w}|}{v_{F}}\right)^{2}\frac{2\pi^{3}\left(-3\cos^{2}(\theta)\left(4k_{F}^{2}\mathrm{R}^{2}+5\right)+k_{F}\mathrm{R}\sin(2k_{F}\mathrm{R})\left(\cos(2\theta)\left(2k_{F}^{2}\mathrm{R}^{2}+15\right)-2k_{F}^{2}\mathrm{R}^{2}+9\right)\right)}{\mathrm{R}^{5}}\nonumber\\
 & +\left(\frac{|\mathbf{w}|}{v_{F}}\right)^{2}\frac{2\pi^{3}\left(4k_{F}^{2}\mathrm{R}^{2}+\cos(2\text{\ensuremath{k_{F}}}\mathrm{R})\left(2k_{F}^{2}\mathrm{R}^{2}+\cos^{2}(\theta)\left(-8k_{F}^{4}\mathrm{R}^{4}-18k_{F}^{2}\mathrm{R}^{2}+15\right)-3\right)+3\right)}{\mathrm{R}^{5}}.
\end{align}
In the limit of large $\mathrm{R}$, the functions $\chi^{(0,\eta)}\left(\mathrm{\mathrm{R}}\right)\propto\cos(2k_{F}\mathrm{R})/\mathrm{\mathrm{R}}^{3}$
and $\left(\partial/\partial\mathrm{R}\right)^{2}\chi^{(2,\eta)}\left(\mathrm{\mathrm{R}}\right)\propto2/\mathrm{R}-\cos(2k_{F}\mathrm{R})/\mathrm{R}$.
The second term in Eq.(\ref{eq: intra-total-range}) provides two
types of coupling: (i) the tilt direction ($\mathbf{w}$) with the
impurity separation direction ($\mathbf{R}$), and (ii) the coupling
between $\mathbf{Q}$ and $\mathbf{R}$ from the $\cos(2\mathbf{Q}\cdot\mathbf{R})$
term. The $\cos(2\mathbf{Q}\cdot\mathbf{R})$ term has already
been studied by the Refs. \cite{PhysRevB.99.165111}. However the
effect of that term would be negligible in two special scenarios,:
(i) in the limit of large Weyl node separation ($|\mathbf{Q}|\gg\mathrm{k_{F}}$),
in which case the summation over all the impurities i.e. summation
over all $\mathbf{R}$ becomes negligibly small because of the very
high oscillation of the cosine function and as a consequence the term
does not result in any significant directional dependence; (ii) in
the limit of small Weyl node separation ($|\mathbf{Q}|\ll\mathrm{k_{F}}$)
in which case the envelope function varies slowly over $\mathrm{R}$
and as a consequence the range function is mostly determined by the
usual RKKY function within the envelope. 

The above results show that the RKKY range function exhibits a distinct
anisotropy in the presence of tilt vector $\mathbf{w}$. This is unlike
the RKKY range function in  graphene, Rashba or other spintronics
systems, where the RKKY is isotropic and dependent only on the distance
between the impurities ($\mathrm{R}$). We will now analyze the anisotropy
of the RKKY range function numerically.

\begin{figure}
\begin{centering}
\subfloat[]{\includegraphics[width=0.5\linewidth]{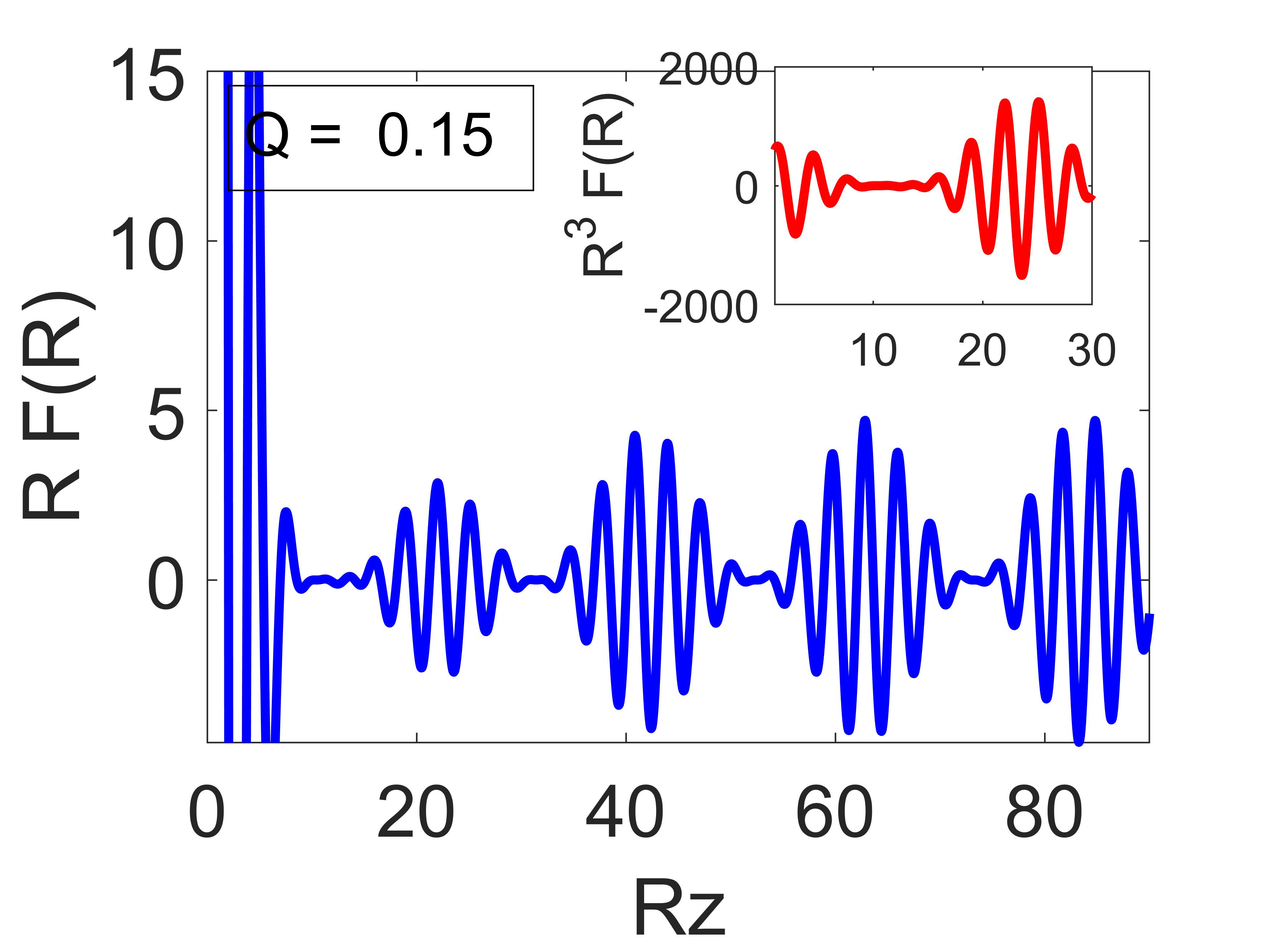}}
\subfloat[]{\includegraphics[width=0.5\linewidth]{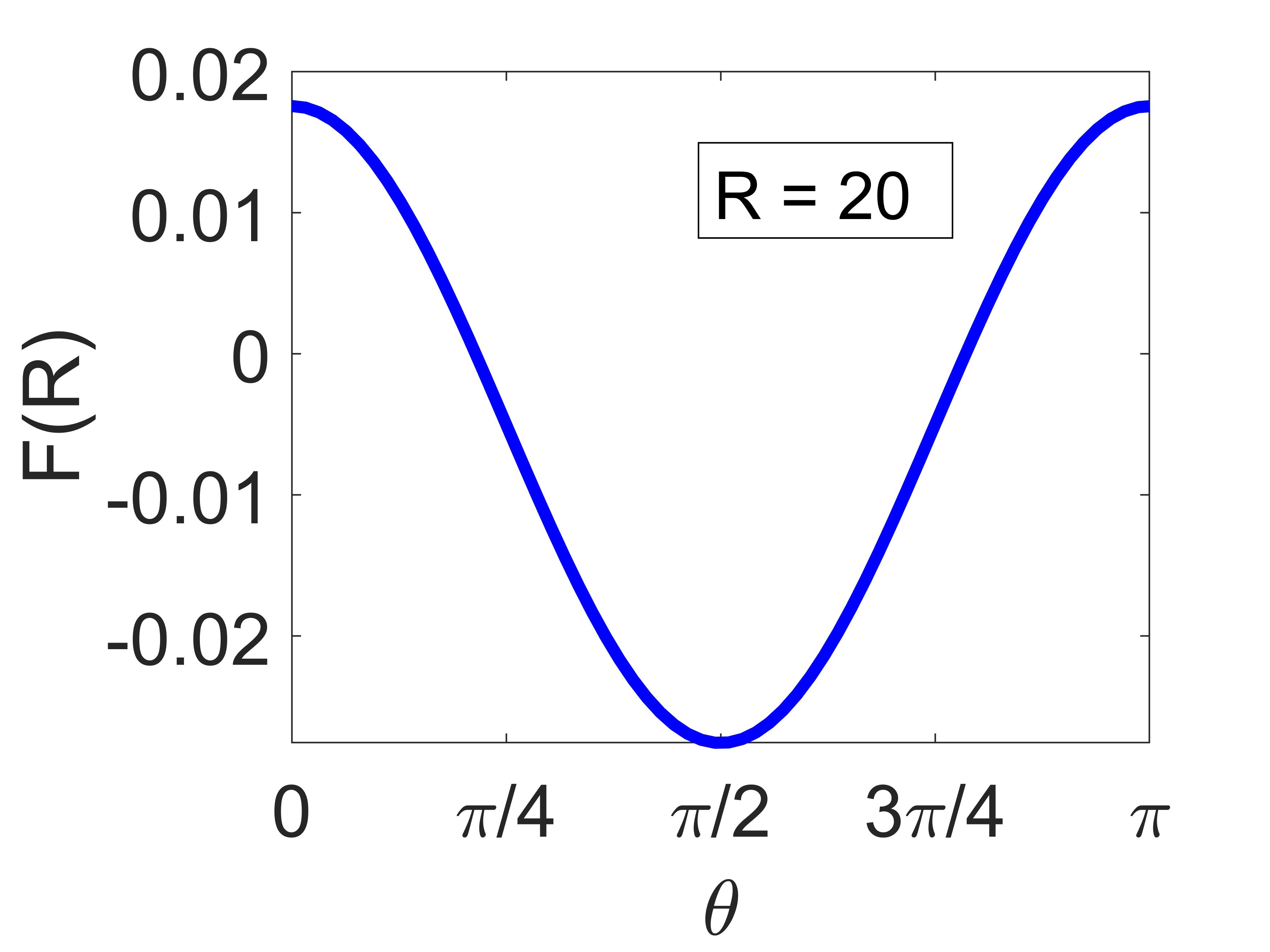}}
\vfill{}
\subfloat[]{\includegraphics[width=0.5\linewidth]{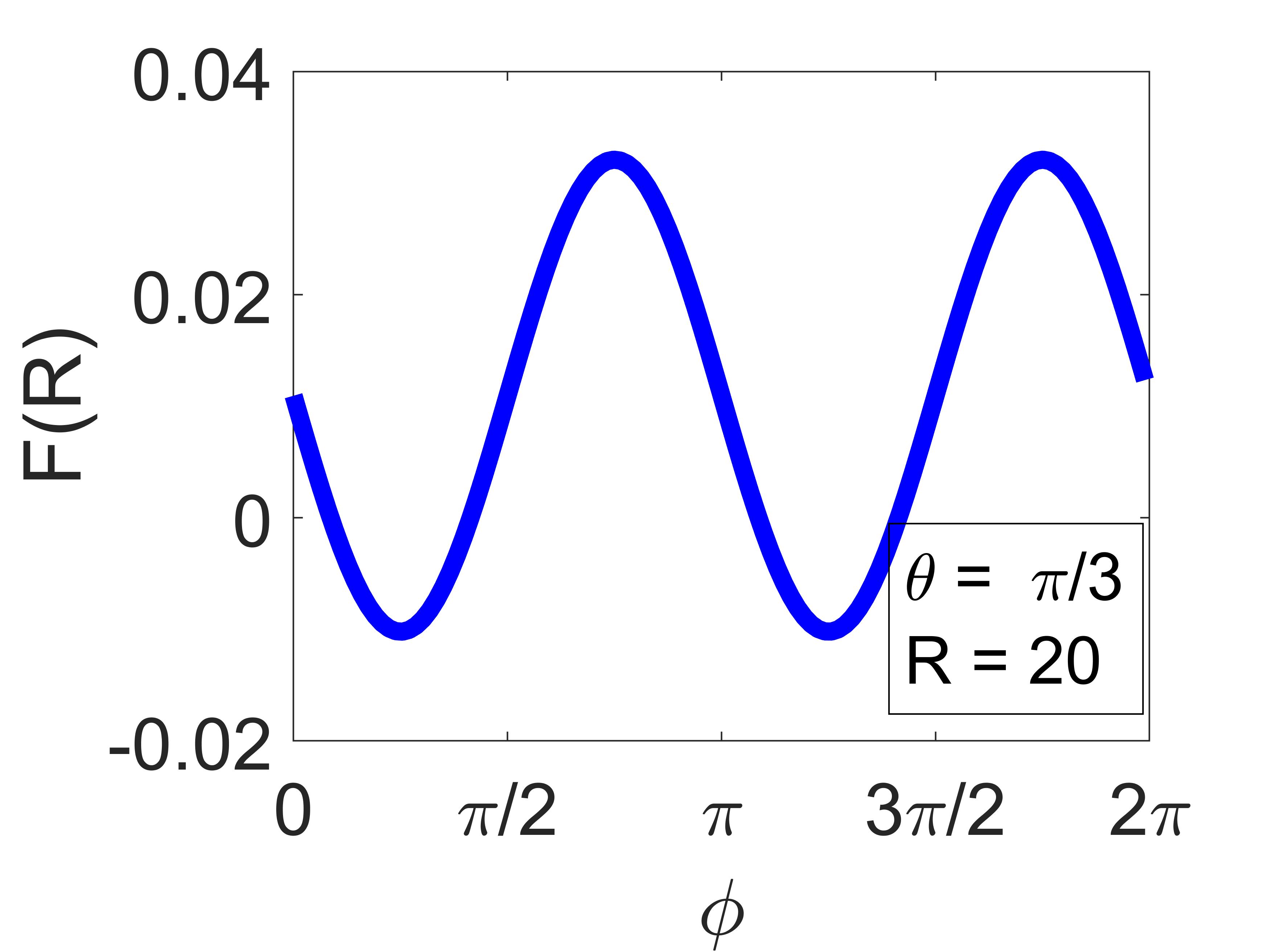}}
\subfloat[]{\includegraphics[width=0.5\linewidth]{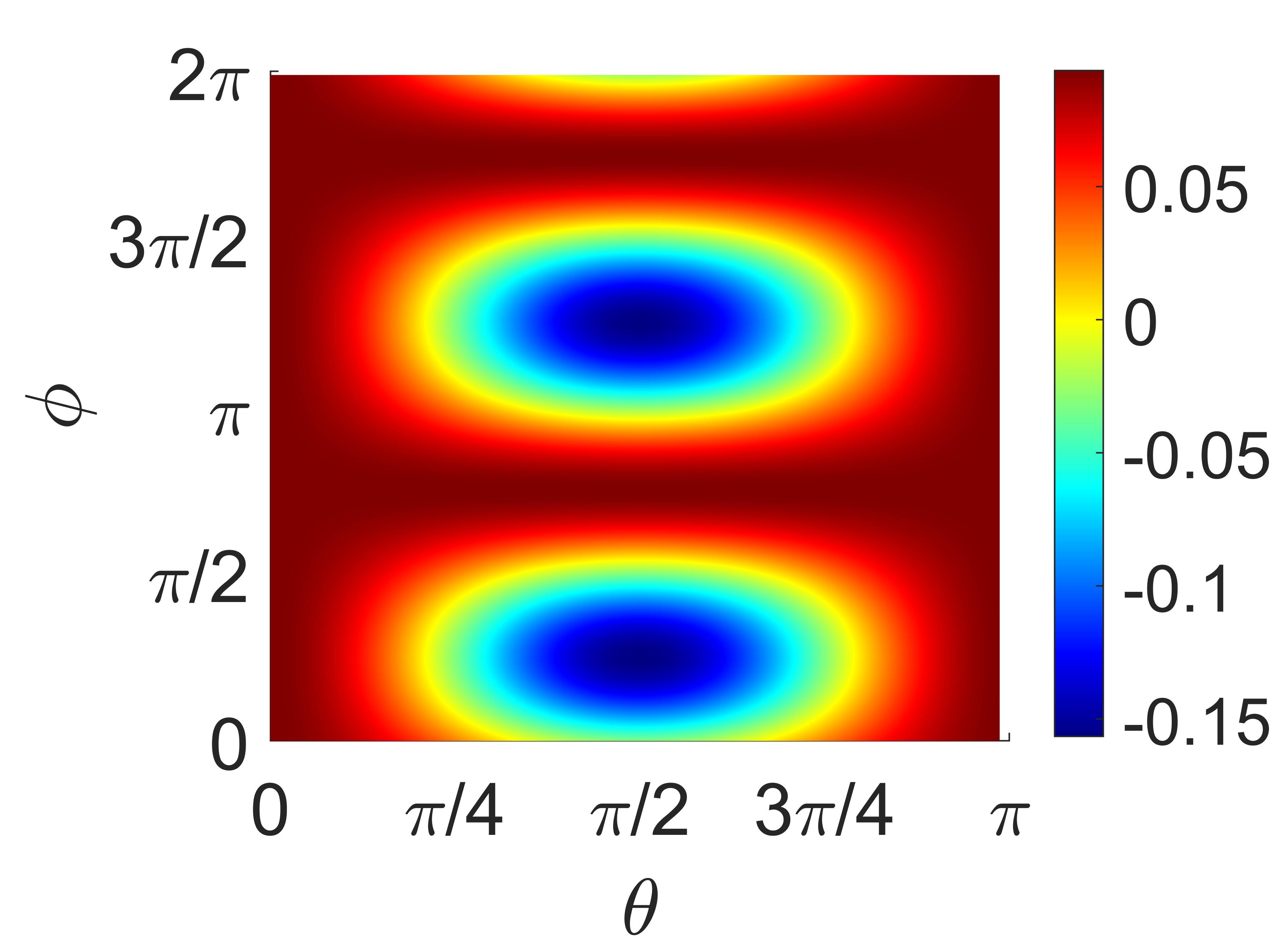}}
\par\end{centering}
\caption{\label{Fig: trsanalytical} WSMs with broken TRS: $|\mathbf{w}|/v_{F}=0.05$
(a) $\mathrm{R}F(\mathbf{R})$ is plotted as a function of $\mathrm{R}$,
for the tilt $\mathbf{w}\parallel\hat{z}.$ In the inset, $\mathrm{R}^{3}F(\mathbf{R})$
is plotted as a function of $\mathrm{R}$. (b) $F(\mathbf{R})$ is
plotted as a function of $\theta$ for the tilt vector $\mathbf{w}\parallel\hat{z}$.
(c) $F(\mathbf{R})$ is plotted as a function of $\phi$ and for the
tilt vector $\mathbf{w}\parallel\hat{x}+\hat{y}$ (d) $F(\mathbf{R})$
is plotted as a function of both polar angle $\theta$ and azimuthal
angle $\phi$. For (b)-(d), we set $k_{F}\mathrm{R}=20$.}
\end{figure}

\textit{Numerical results for broken TRS:} In our numerical calculations,
we set $\mathbf{R}=\mathrm{R}(\sin(\theta)\cos(\phi)\hat{x}+\sin(\theta)\sin(\phi)\hat{y}+\cos(\theta)\hat{z})$.
In a WSM the $x$, $y$, $z$-axes may coincide with the crystal axes.
The integrations in Eqs. (\ref{eq:Fintramain}) and (\ref{eq:Fintermain})
are performed numerically for $\epsilon_{0}=0$. In Fig \ref{Fig: trsanalytical}(a),
we plot the total RKKY range function $F(\mathbf{R})$ while varying
$\mathrm{R}$ for the special case of $\mathbf{w}\parallel\mathbf{Q}\parallel\mathbf{R}\parallel\hat{z}$.
At the small R limit, the envelope function has a $1/\mathrm{R}^{3}$
dependence, as can be seen from the inset. In this limit, the RKKY
function has a similar dependence compared to the non-tilted WSMs.
However for a sufficiently large value of $\mathrm{R}$ and in the
presence of tilt the envelope function approaches $1/\mathrm{R}$
dependence matching our analytical results. Next, in Fig. \ref{Fig: trsanalytical}(b)
and (c), we plot the angular dependence of the range function $F(\mathbf{R})$.
In Fig. \ref{Fig: trsanalytical}(b), we show the angular dependence
of $F(\mathbf{R})$ by varying the angle ($\theta$) between $\mathbf{w}$
and $\mathbf{R}$ while fixing the tilt direction to be $\mathbf{w}\parallel\hat{z}$.
In Fig. \ref{Fig: trsanalytical}(c), we plot the angular dependence
of $F(\mathbf{R})$ by varying the angle ($\phi$) between $\mathbf{R}$
and $x$-axis while fixing the tilt vector $\mathbf{w}\parallel(\hat{x}+\hat{y})$,
i.e., it lies in the $x$-$y$ plane. For both cases, the range function
oscillates with the angle between tilt and the separation vector $\mathbf{R}$.
In Fig \ref{Fig: trsanalytical}(d), we plot the angular dependence
of $F(\mathbf{R})$ as a function of both with polar angle ($\theta$)
and azimuthal angle ($\phi$) while fixing the tilt vector $\mathbf{w}\parallel(\hat{x}+\hat{y})$.
There is significant variation in the range function with both $\theta$
and $\phi$, i.e., showing a strong directional dependence on the
separation vector $\mathbf{R}$.

\subsection{Broken IR ($\mathbf{Q}=0$ and $\epsilon_0\protect\neq0$)} 

Next, we investigate the WSM systems where the inversion symmetry
is broken. The corresponding Hamiltonian can be obtained by letting
$\mathbf{Q}=0$ in the Hamiltonian of Eq. (1), i.e., $H_{0}=\eta v_{F}\boldsymbol{\tau}\cdot\mathbf{k}+\mathbf{w}_{\eta}\cdot\mathbf{k}\tau_{0}+\eta\epsilon_{0}\tau_{0}$.
The Dirac nodes in this case are separated along the energy axis.
In this case, the Fermi surface has different sizes for the two different
valleys, i.e., ${k}_{\mathrm{F}\eta}\neq {k}_{\mathrm{F}\eta}$
and as a consequence, the intra and inter-valley range functions are
different from that of the TRS-breaking case. Assuming parallel tilt
vectors for both the valleys, i.e., $\mathbf{w}_{\eta}=\mathbf{w}$
and using the Eq. (\ref{eq:Fintramain}) and Eq. (\ref{eq:Fintermain});
the expression for the total range function is given as follows, 
\begin{align}
F\left(\mathbf{R}\right) & =\left\{ \sum_{\eta}\chi_{IR}^{(0,\eta)}\left(\mathrm{R}\right)-\sum_{\eta}\left(\frac{\mathbf{w}}{v_{F}}\cdot\frac{\partial}{\partial\mathbf{R}}\right)^{2}\chi_{IR}^{(2,\eta)}\left(\mathrm{R}\right)\right\} \nonumber\\
 & +\left\{ \sum_{\eta}\left[\zeta_{IR}^{(0)}\left(\eta,-\eta,\mathrm{R}\right)-\left(\frac{\mathbf{w}}{v_{F}}\cdot\frac{\partial}{\partial\mathbf{R}}\right)^{2}\zeta_{IR}^{(2)}\left(\eta,-\eta,\mathrm{R}\right)\right]\right\} .
\end{align}
In the above, the intra-valley and the inter-valley contributions
at zero tilt are given explicitly as follows, 
\begin{align}
\sum_{\eta}\chi_{IR}^{(0,\eta)}\left(\mathrm{R}\right) & =\frac{4\pi^{3}\left(\left(3-6\mathrm{R}^{2}(q_{0}-\text{\ensuremath{k_{F}}})^{2}\right)\cos(2\mathrm{R}(\text{\ensuremath{k_{F}}}-q_{0}))+\left(3-6\mathrm{R}^{2}(q_{0}+\text{\ensuremath{k_{F}}})^{2}\right)\cos(2\mathrm{R}(q_{0}+\text{\ensuremath{k_{F}}}))\right)}{3\mathrm{R}^{5}}\nonumber\\
 & +\frac{4\pi^{3}\left(6\mathrm{R}(\text{\ensuremath{k_{F}}}-q_{0})\sin(2\mathrm{R}(\text{\ensuremath{k_{F}}}-q_{0}))+6\mathrm{R}(q_{0}+\text{\ensuremath{k_{F}}})\sin(2\mathrm{R}(q_{0}+\text{\ensuremath{k_{F}}}))-6\right)}{3\mathrm{R}^{5}}
, \end{align}
 \begin{align}
\sum_{\eta}\zeta_{IR}^{(0)}\left(\eta,-\eta,\mathrm{R}\right) & =\frac{4\pi^{3}\left(-4q_{0}\mathrm{R}\left(4\mathrm{R}^{2}\left(2q_{0}^{2}+3\text{\ensuremath{k_{F}}}^{2}\right)+3\right)\sin(2q_{0}\mathrm{R})\right)}{3\mathrm{R}^{5}}\nonumber\\
 & +\frac{4\pi^{3}\left(3\left(1-2\mathrm{R}^{2}\left(3q_{0}^{2}-4q_{0}\text{\ensuremath{k_{F}}}+\text{\ensuremath{k_{F}}}^{2}\right)\right)\cos(2\mathrm{R}(\text{\ensuremath{k_{F}}}-2q_{0}))\right)}{3\mathrm{R}^{5}}\nonumber\\
 & +\frac{4\pi^{3}\left(-3\left(2\mathrm{R}^{2}(q_{0}+\text{\ensuremath{k_{F}}})(3q_{0}+\text{\ensuremath{k_{F}}})-1\right)\cos(2\mathrm{R}(2q_{0}+\text{\ensuremath{k_{F}}}))\right)}{3\mathrm{R}^{5}}\nonumber\\
 & +\frac{4\pi^{3}\left(6\mathrm{R}(\text{\ensuremath{k_{F}}}-2q_{0})\sin(2\mathrm{R}(\text{\ensuremath{k_{F}}}-2q_{0}))+6\mathrm{R}(2q_{0}+\text{\ensuremath{k_{F}}})\sin(2\mathrm{R}(2q_{0}+\text{\ensuremath{k_{F}}}))-6\cos(2q_{0}\mathrm{R})\right)}{3\mathrm{R}^{5}}
,\end{align}
and, the intra-valley and the inter-valley contributions to the second
order in the tilt parameter can be expressed as follows, 
\begin{align}
\chi_{IR}^{(2,\eta)}\left(\mathrm{R}\right) & =\frac{4\pi^{3}\left(-4\mathrm{R}^{2}(q_{F}-\eta q_{0})^{2}+\left(2\mathrm{R}^{2}(q_{F}-\eta q_{0})^{2}+1\right)\cos(2\mathrm{R}(q_{F}-\eta q_{0}))+2\mathrm{R}(q_{F}-\eta q_{0})\sin(2\mathrm{R}(q_{F}-\eta q_{0}))-1\right)}{\mathrm{R}^{3}}
, \end{align}
 \begin{align}
\zeta_{IR}^{(2)}\left(\eta,-\eta,\mathrm{R}\right) & =\frac{4\pi^{3}\left(2\mathrm{R}\left(3\eta q_{0}-2\mathrm{R}^{2}(k_{F}-4\eta q_{0})(k_{F}-\eta q_{0})^{2}\right)\sin(2\eta q_{0}\mathrm{R})-3\left(4\mathrm{R}^{2}(k_{F}-\eta q_{0})^{2}+1\right)\cos(2\eta q_{0}\mathrm{R})\right)}{3\mathrm{R}^{3}}\nonumber\\
 & +\frac{4\pi^{3}\left(3\left(2\mathrm{R}^{2}(k_{F}-\eta q_{0})(k_{F}-3\eta q_{0})+1\right)\cos(2\mathrm{R}(k_{F}-2\eta q_{0}))+6k_{F}\mathrm{R}\sin(2\mathrm{R}(k_{F}-2\eta q_{0}))\right)}{3\mathrm{R}^{3}}
,\end{align}
where, $q_{0}=\epsilon_{0}/\hbar v_{F}$. The full expression of all
the second-order derivative functions are given in the Supplemental.
Similar to the TRS case the $\chi^{(0,\eta)}\left(\mathrm{R}\right)$
and $\zeta^{(0)}\left(\eta,-\eta,\mathrm{\mathrm{R}}\right)$ functions
have a $1/\mathrm{R}^{3}$ dependence while $\chi^{(2,\eta)}\left(\mathrm{R}\right)$
varies with $1/\mathrm{R}$. Specifically at large $R$, the 1/R dependence
dominates and the total range function varies as,
\begin{align}
F\left(\mathbf{R}\right) & \propto\frac{16\pi^{3}|\mathbf{w}|^{2}}{\mathrm{R}}\nonumber \\
 & \left(-4k_{F}^{2}q_{0}^{2}\cos(2q_{0}\mathrm{R})+2k_{F}^{2}(k_{F}^{2}-q_{0}^{2})\cos(2k_{F}\mathrm{R})+(k_{F}-q_{0})^{4}\cos(2\mathrm{R}(k_{F}-q_{0}))+(k_{F}+q_{0})^{4}\cos(2\mathrm{R}(k_{F}+q_{0}))\right).
\end{align}

\begin{figure}
\begin{centering}
\subfloat[]{\includegraphics[width=0.5\linewidth]{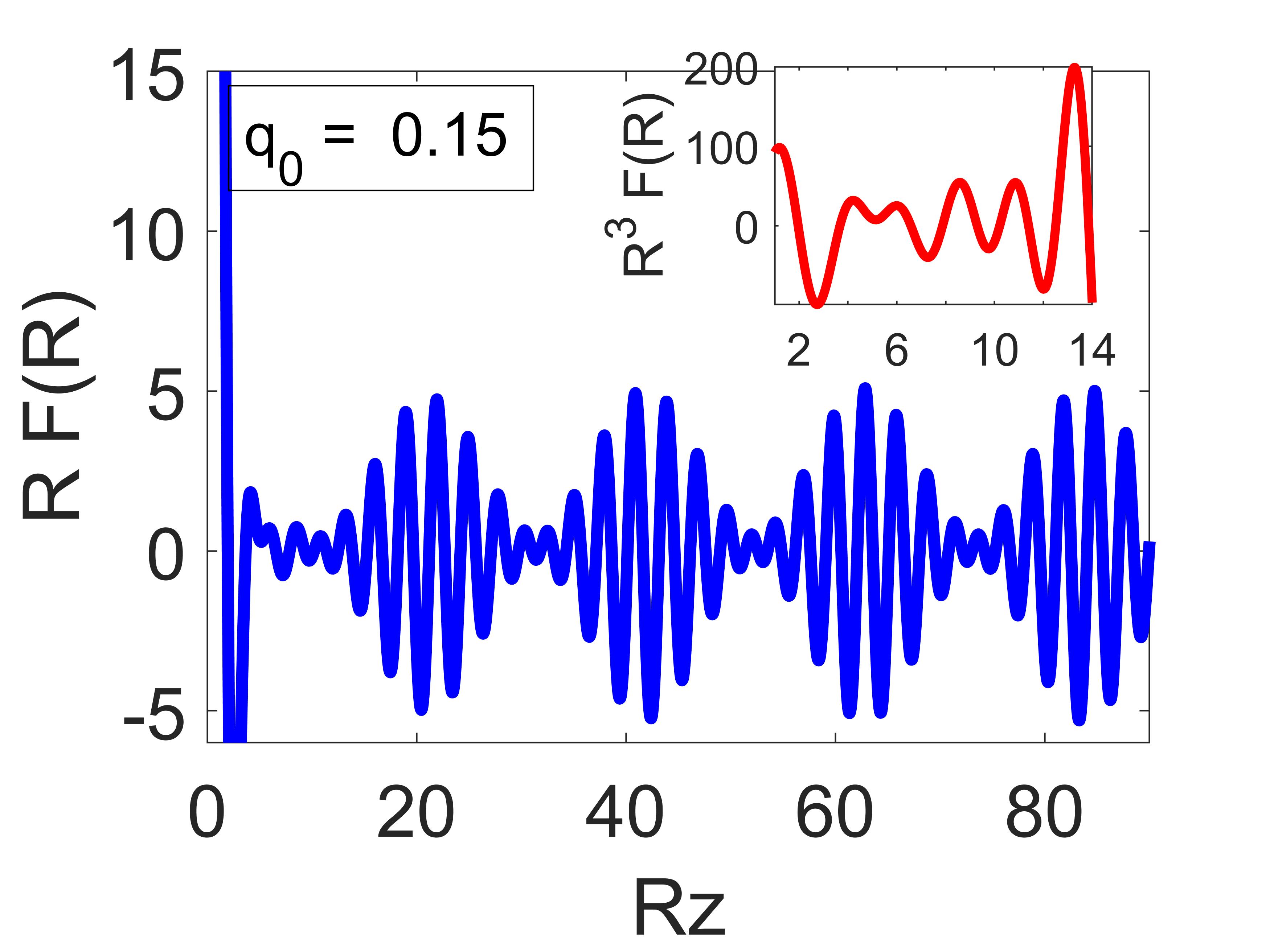}}
\subfloat[]{\includegraphics[width=0.5\linewidth]{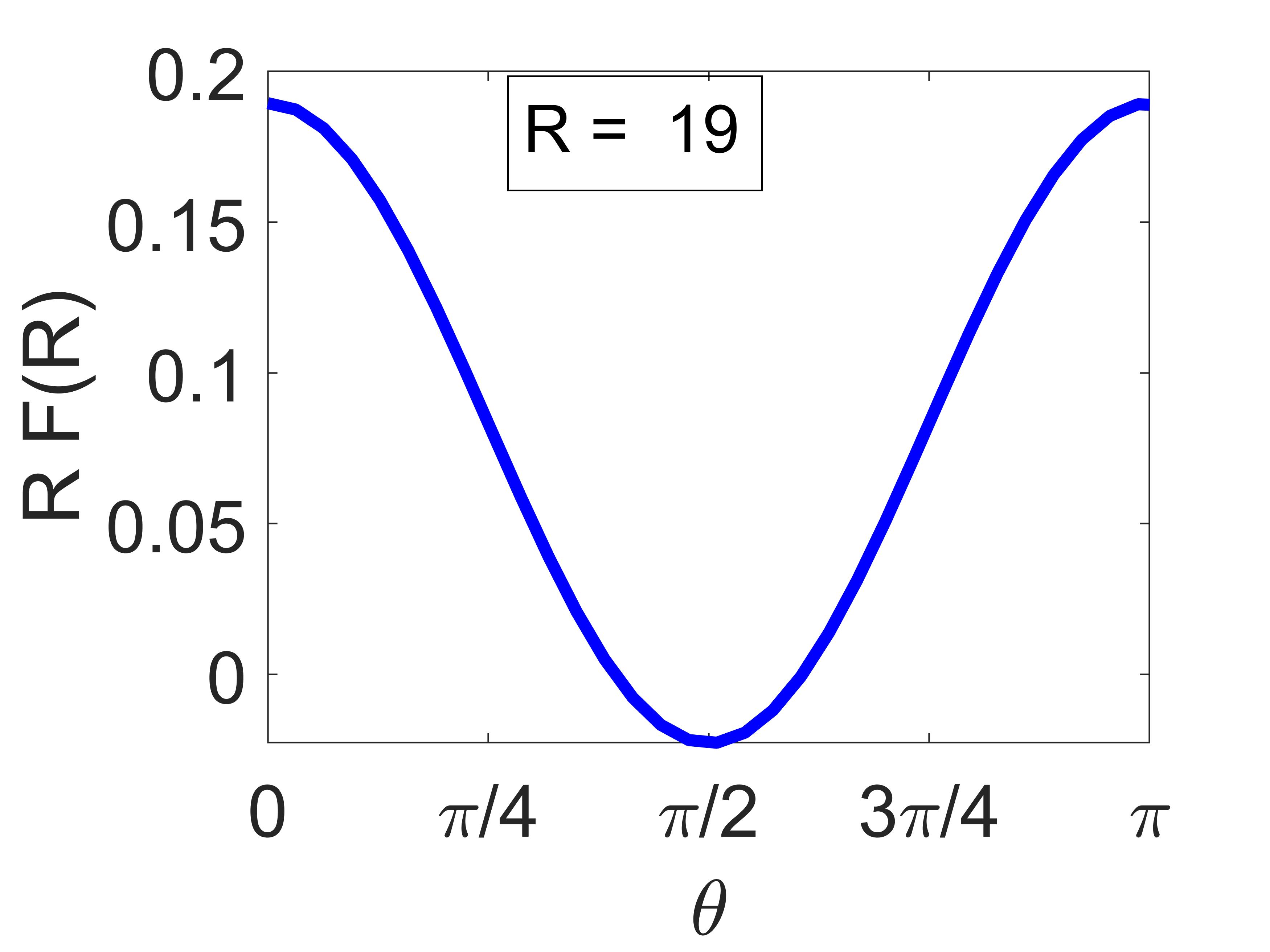}}
\vfill{}
\subfloat[]{\includegraphics[width=0.5\linewidth]{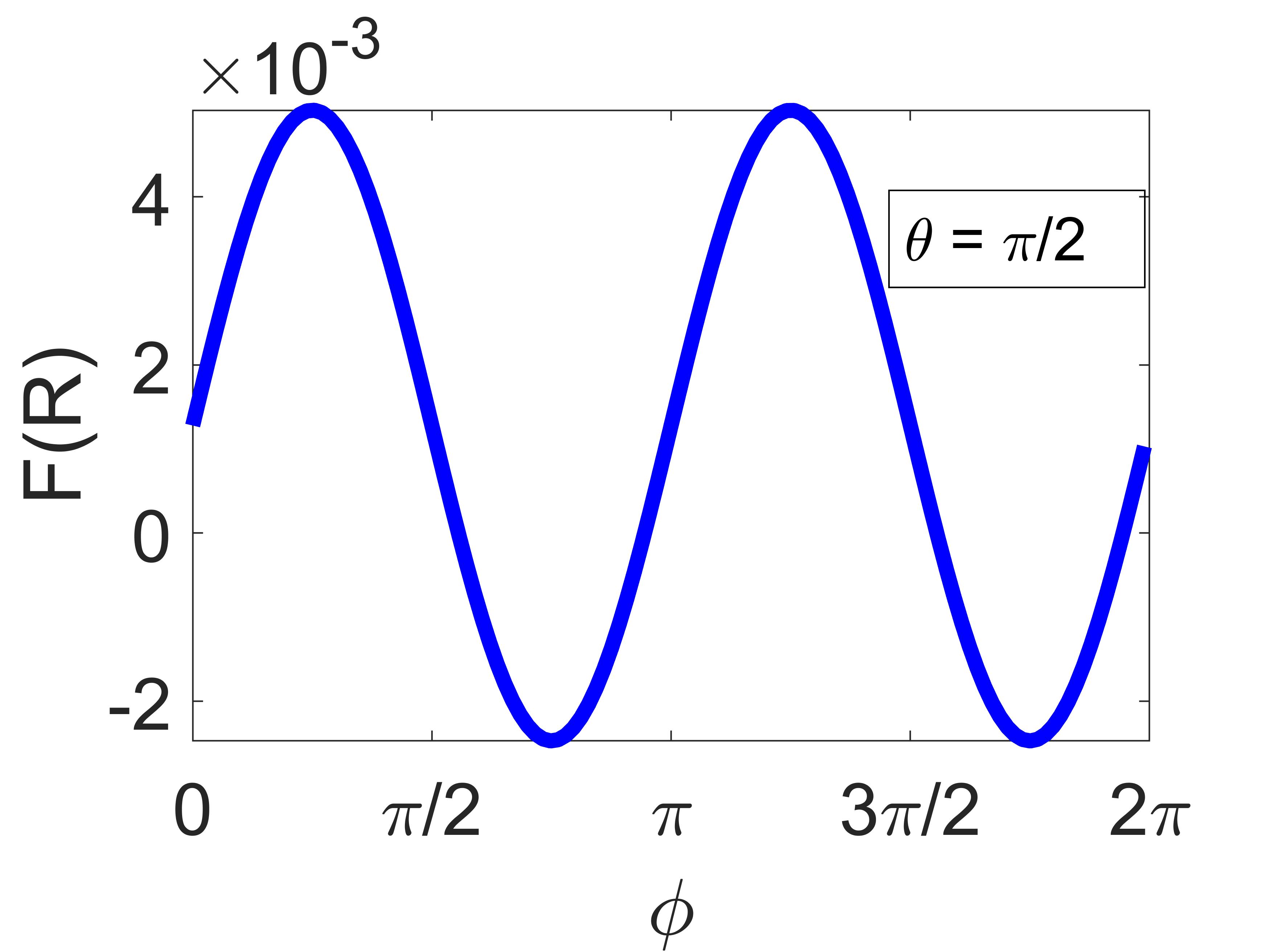}}
\subfloat[]{\includegraphics[width=0.5\linewidth]{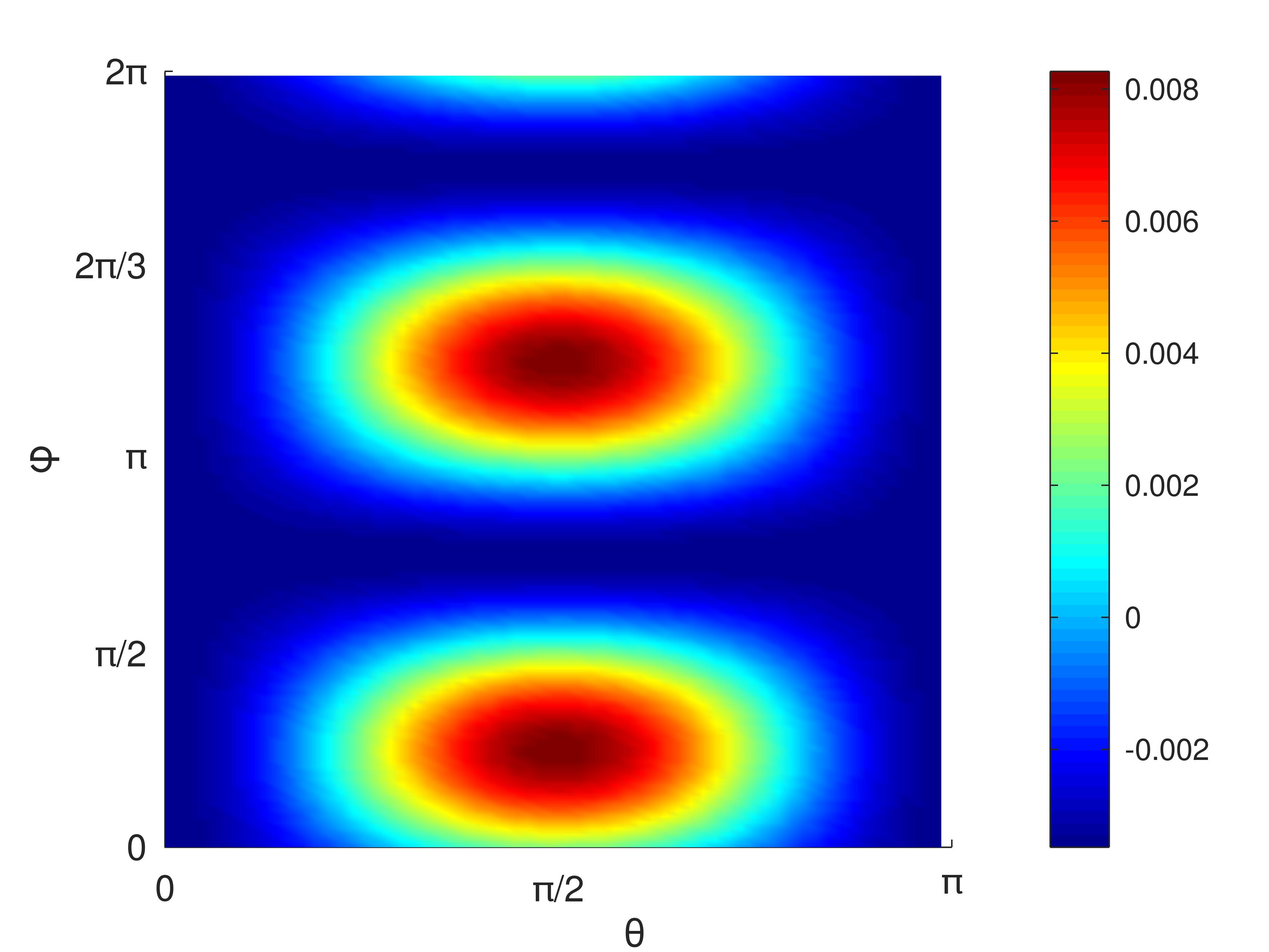}}
\par\end{centering}
\begin{centering}
\vfill{}
\par\end{centering}
\caption{\label{Fig: iranalytical} IR breaking WSM: $|\mathbf{w}|/v_{F}=0.05$,
(a) Plot of $\mathrm{R}F(\mathrm{R})$ and the inset is the small
$\mathrm{R}$ limit, for the tilt $\mathbf{w}\parallel\hat{z}$ (b)
Range function vs. $\theta$ for the tilt, $\mathbf{w}\parallel\hat{z}$
(c) Range function vs. $\phi$ and for tilt, $\mathbf{w}\parallel\hat{x}+\hat{y}$
(d) Range function vs. both polar angle $\theta$ and azimuthal angle
$\phi$. For (b)-(d), we choose $k_{F}\mathrm{R}=20$.}
\end{figure}

\textit{Numerical results for broken IR:} The integrations in Eqs. (\ref{eq:Fintramain}) and (\ref{eq:Fintermain})
are performed numerically for $\bf{Q}=0$. In Fig \ref{Fig: iranalytical}(a),
we plot the total RKKY range function $\mathrm{R}F(\mathbf{R})$ while
varying $\mathrm{R}$ for the special case of $\mathbf{w}\parallel\mathbf{R}\parallel\hat{z}$.
At the small $\mathrm{R}$ limit, the envelope function varies with
$1/\mathrm{R}^{3}$ as can be seen from the inset. In this limit,
the RKKY function has a similar dependence as that of non-tilted WSMs.
However for a sufficiently large value of $\mathrm{R}$ and in the
presence of tilt the envelope function has a $1/\mathrm{R}$ dependence.
In Fig. \ref{Fig: iranalytical}(b), we show that the angular dependence
of $F(\mathbf{R})$ by varying the angle between $\mathbf{w}$ and
$\mathbf{R}$ while fixing the tilt direction $\mathbf{w}\parallel\hat{z}$.
In Fig \ref{Fig: iranalytical}(c), we plot the angular dependence
of $F(\mathbf{R})$ by varying the angle between $\mathbf{R}$ and
$x$-axis while fixing the tilt vector $\mathbf{w}\parallel(\hat{x}+\hat{y})$
i.e. tilt vector lies in the $x$-$y$ plane. In Fig \ref{Fig: iranalytical}(d),
we show the angular dependence of $F(\mathbf{R})$ as a function of
both the polar angle ($\theta$) and azimuthal angle ($\phi$) while
fixing the tilt vector $\mathbf{w}\parallel(\hat{x}+\hat{y})$. For
Fig \ref{Fig: iranalytical}(a)-(d) we choose $q_{0}/k_{F}=0.15$.

Several observations can be made immediately by comparing Fig. \ref{Fig: iranalytical}
and Fig. \ref{Fig: trsanalytical}. The separation between the nodes
i.e. $|\mathbf{Q}|$ for TRS breaking and $q_{0}$ for the IR breaking
WSMs respectively produce similar envelope function in both Fig. \ref{Fig: trsanalytical}(a)
and Fig. \ref{Fig: iranalytical}(a). The difference in magnitude
between TRS and IR-breaking system is not very significant. For both
the TRS and IR-breaking WSM systems, it is possible to switch the
RKKY-induced exchange coupling from ferromagnetic (i.e., $F(\mathbf{R})<0$)
to anti-ferromagnetic (i.e., $F(\mathbf{R})>0$) coupling by varying
both the angles $\theta$ and $\phi$ associated with the separation
vector $\mathbf{R}$, as shown in Figs. \ref{Fig: trsanalytical}(d)
and \ref{Fig: iranalytical}(d).

Note that for the TRS breaking case, the directional dependence of
the range function arises due to the $\cos(2\mathbf{Q}\cdot\mathbf{R})$
term as can be seen from Eq. (\ref{eq: intra-total-range}). This
term produces an extra oscillation with respect to the angular variation
on top of the intra-valley range function \cite{PhysRevB.92.224435}.
However the effect of this term vanishes for the special case when
$\mathbf{Q}\perp\mathbf{R}$, i.e., when the separation between nodes
is perpendicular to spatial vector between impurities. For the case
of IR breaking WSMs, the analytical expression of the envelope function
cannot be factored out analytically, and thus we cannot similarly
find a spatial direction where the oscillatory dependence on the angular
direction vanishes.

\subsection{Opposite Tilt\textit{ ($\mathbf{w}_{\eta}=\eta\mathbf{w}$)}}

\begin{figure}
\begin{centering}
\subfloat[]{\includegraphics[width=0.5\linewidth]{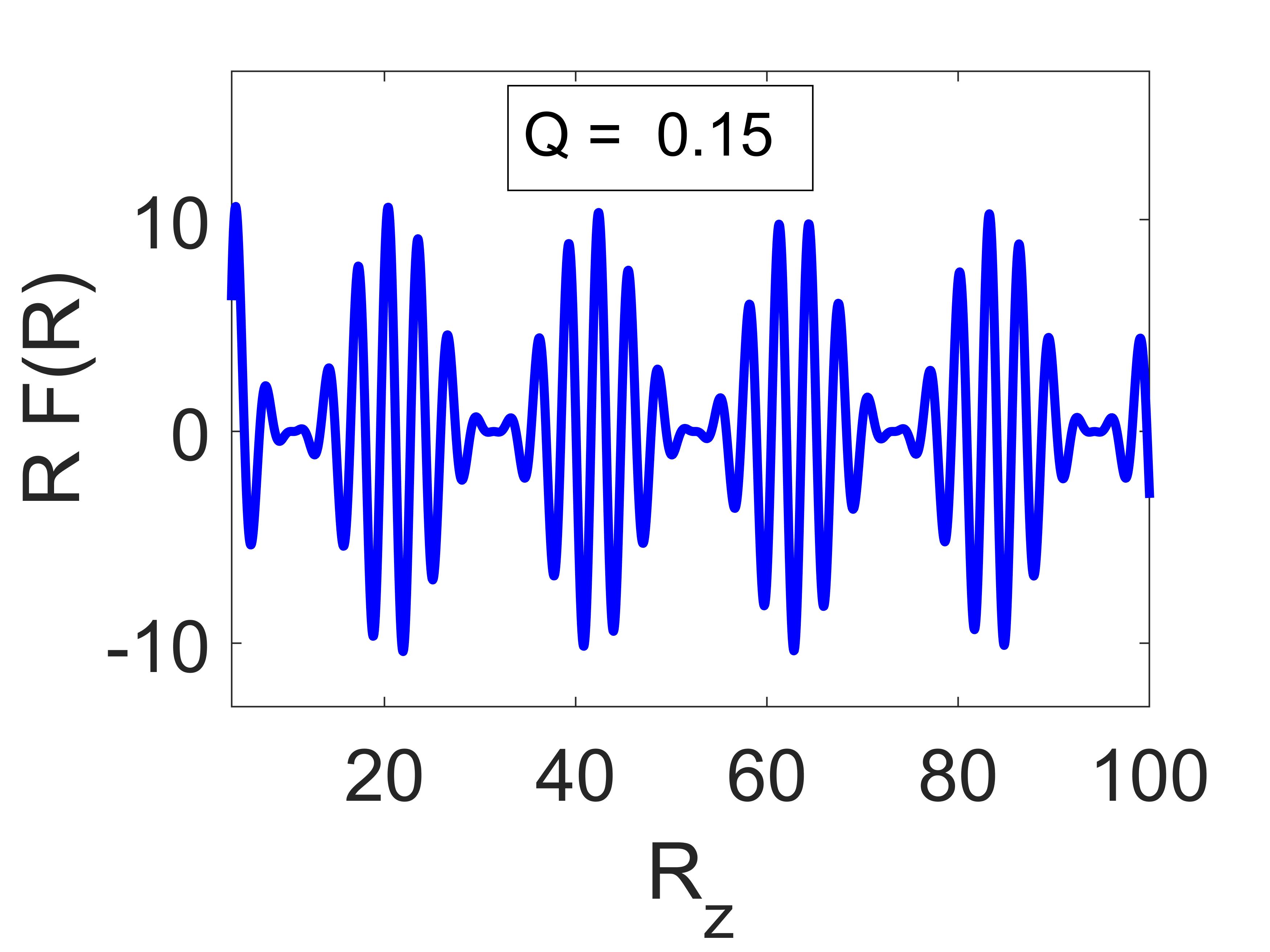}}
\subfloat[]{\includegraphics[width=0.5\linewidth]{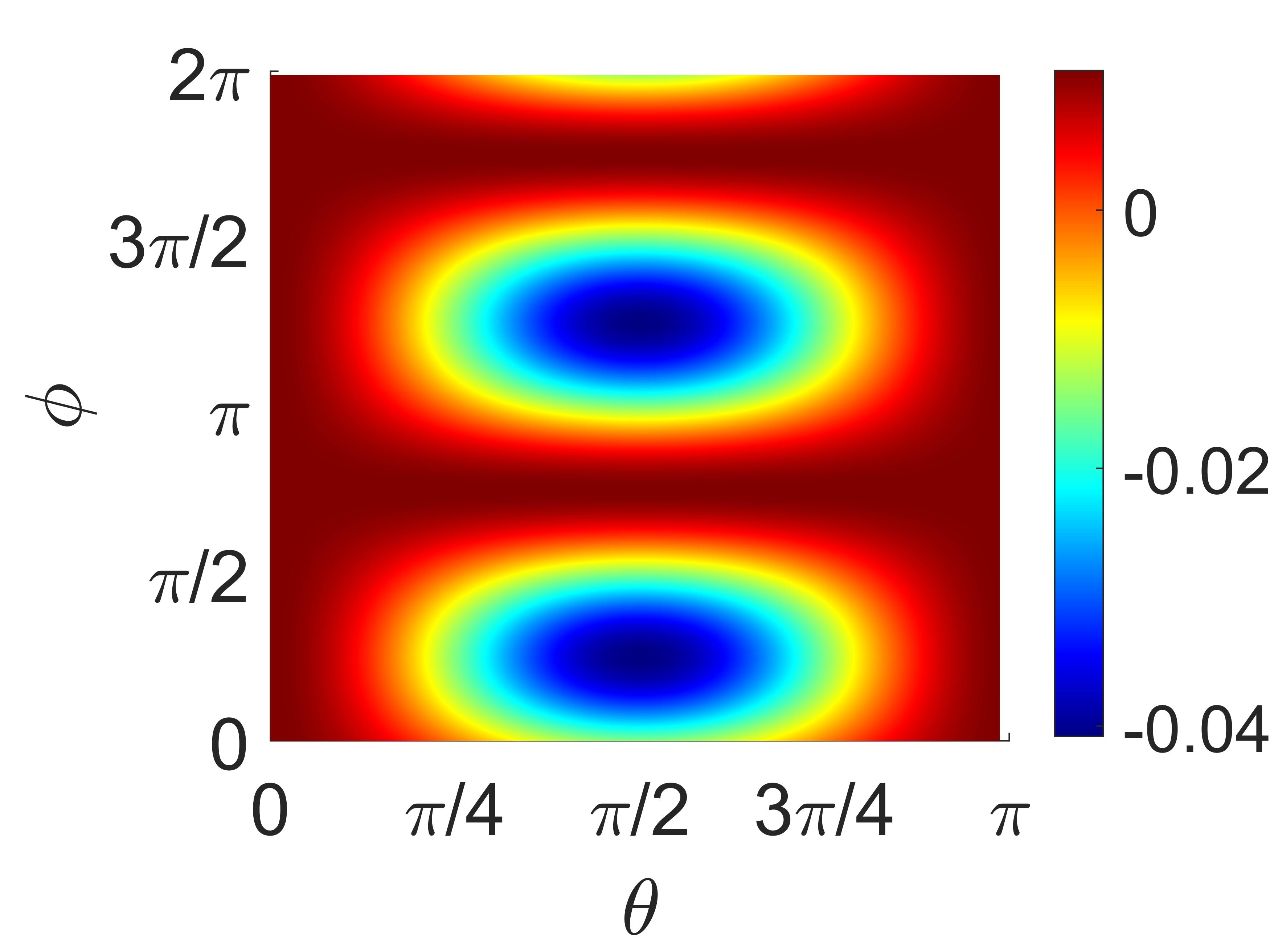}}
\par\end{centering}
\caption{\label{Fig: trsanalytical-1} TRS breaking WSM with the tilt $\mathbf{w}\parallel\hat{z}$
and $|\mathbf{w}|/v_{F}=0.05$: (a) Plot of $\mathrm{R}F(\mathrm{R})$
as a function of $\mathrm{R}$ which is along the $z$-direction,
(b) Contour plot of the range function $F(\mathrm{R})$ as a function
of both polar angle $\theta$ and azimuthal angle $\phi$ with $k_{F}\mathrm{R}=20$.}
\end{figure}

We also calculate the RKKY range function for the case where the dispersion
cones are tilted in opposite directions for Weyl nodes of opposite
chirality (i.e. $\mathbf{w}_{\eta}=\eta\mathbf{w}$). In this case,
the range function differs from the parallel tilt case in the second-order
in terms of the tilt parameter. 
\textbf{\textit{TRS broken ($\epsilon_{0}=0$):}} The total range
function is now given by, 
\begin{align}
F\left(\mathbf{R}\right)=2\left(1+\cos\left(2\mathbf{Q}\cdot\mathbf{R}\right)\right)\left(\chi^{(0,\eta)}\left(\mathrm{R}\right)+\chi^{(2,\eta)}\left(\mathrm{R}\right)\right),
\end{align}
where,
\begin{align}
\chi^{(2,\eta)}\left(\mathrm{R}\right) & =-\text{Re}\int_{k=0}^{k_{F\eta}}d\boldsymbol{k}\text{ }\int_{k'=0}^{\infty}d\boldsymbol{k'}\frac{\left(\boldsymbol{\mathrm{w}}\cdot\left(\boldsymbol{k'}+\boldsymbol{k}\right)\right)^{2}e^{i\left(\boldsymbol{k}'-\boldsymbol{k}\right)\cdot\mathbf{R}}}{\left(\hbar v_{F}\left(k'-k\right)\right)^{3}}
.\end{align}
We rewrite the above equation as,
\textbf{\textit{ 
\begin{align}
\chi^{(2,\eta)}\left(\mathrm{R}\right)
 & =\text{Re}\left(\left(\frac{\boldsymbol{\mathbf{w}}}{v_{F}}\cdot\frac{\partial}{\partial\mathbf{R}}\right)^{2}\int_{k=0}^{k_{F\eta}}d\boldsymbol{k}\text{ }e^{-i\boldsymbol{k}\cdot\mathbf{R}}\right)\int_{k'=0}^{\infty}d\boldsymbol{k'}\frac{e^{i\boldsymbol{k}'\cdot\mathbf{R}}}{\left(\hbar v_{F}\left(k'-k\right)\right)^{3}}\nonumber\\
 & +\text{Re}\int_{k=0}^{k_{F\eta}}d\boldsymbol{k}\text{ }e^{-i\boldsymbol{k}\cdot\mathbf{R}}\left[\left(\frac{\boldsymbol{\mathbf{w}}}{v_{F}}\cdot\frac{\partial}{\partial\mathbf{R}}\right)^{2}\int_{k'=0}^{\infty}d\boldsymbol{k'}\frac{e^{i\boldsymbol{k}'\cdot\mathbf{R}}}{\left(\hbar v_{F}\left(k'-k\right)\right)^{3}}\right]\nonumber\\
 & +2\text{Re}\left[\left(\frac{\boldsymbol{\mathbf{w}}}{v_{F}}\cdot\frac{\partial}{\partial\mathbf{R}}\right)\int_{k=0}^{k_{F\eta}}d\boldsymbol{k}\text{ }e^{-i\boldsymbol{k}\cdot\mathbf{R}}\right]\left[\left(\frac{\boldsymbol{\mathbf{w}}}{v_{F}}\cdot\frac{\partial}{\partial\mathbf{R}}\right)\int_{k'=0}^{\infty}d\boldsymbol{k'}\frac{e^{i\boldsymbol{k}'\cdot\mathbf{R}}}{\left(\hbar v_{F}\left(k'-k\right)\right)^{3}}\right]
.\end{align}
}}As before, the full expressions of the above are given in the supplemental
due to the size of the expressions. In Fig \ref{Fig: trsanalytical-1}(a),
we plot the total RKKY range function $F(\mathbf{R})$ while varying
$\mathrm{R}$ for the special case of $\mathbf{w}\parallel\mathbf{Q}\parallel\mathbf{R}\parallel\hat{z}$.
In Fig. \ref{Fig: trsanalytical-1}(b), we plot the angular dependence
of $F(\mathbf{R})$ as a function of both with polar angle ($\theta$)
and azimuthal angle ($\phi$) while we keep $\mathbf{w}\parallel\hat{z}$.
Similar to the parallel tilt case, range function varies with the
changes in $\theta$ and $\phi$, i.e., showing a strong directional
dependence on the separation vector $\mathbf{R}$.

\textit{Finite tilt limit:} Thus far, we have considered the analytical
expression of the RKKY range function in the small-tilt limit. We
also analyze the RKKY range function numerically for any arbitrary
tilt value. In doing so, we consider the general expressions in Eqs.
(\ref{eq: total range function}-\ref{eq:Fintermain}) , assume parallel
tilt vector for all Weyl nodes and set the tilt as $\mathbf{w}\parallel\hat{z}$.
The corresponding plots for these numerical calculations are given
in the supplemental. Generally, in the small tilt limit the RKKY function
is linearly dependent on tilt matching our analytical result. However,
at the large tilt limit, the variation of the RKKY range function
with the tilt magnitude diverges from the analytical results, and
assumes a non-linear variation. 

\section{Discussion}

We explicitly study the effect of tilted energy dispersion on the
RKKY coupling strength for two types of WSMs, namely, the TRS-broken
and IR-broken. The direction of the tilt can be determined from the
band structure and can vary with the crystal axis in different materials.
We fix the direction of tilt and analyze the RKKY range function with
varying magnitude and direction of $\mathbf{R}$. The range function
in the absence of tilt varies as $1/\mathrm{R}^{3}$ for both the
TRS and IR breaking system. Interestingly, however, in the presence
of tilt, the the total range function varies as $1/\mathrm{R}$ in
the limit of large $\mathrm{R}$. This means that the dispersion tilt
can have a significant effect on the overall RKKY effect in WSMs even
though the tilt-dependent term is only second-order in the tilt parameter.
Due to the $1/\mathrm{R}$ dependence, this second-order tilt-dependent
term would still dominate over the tilt-independent term in the limit
of large $\mathrm{R}$. As a consequence, the long range RKKY coupling
in tilted WSMs would be determined primarily by the tilt-induced term.

In both Fig. \ref{Fig: trsanalytical}(a) and \ref{Fig: iranalytical}(a),
for $|\mathbf{w}|=0.05$ the tilt induced $1/\mathrm{R}$ term overcomes
the conventional tilt-independent term (which is proportional to $1/\mathrm{R}^{3}$)
at around $k_{F}\mathrm{R}=40$. If a physical WSM system system has
Fermi energy $\epsilon_{F}$ such that $k_{F}=1 \AA^{-1}$ (i.e.,
$\epsilon_{F}=\hbar^{2}k_{F}^{2}/2m_{e}=3.43$ eV), then $k_{F}\mathrm{R}=40$
corresponds to $\mathrm{R}=4$ nm. This is a feasible distance to
observe the RKKY coupling in practical devices and at this distance
one can expect the effect of dispersion tilt in the RKKY coupling
to become dominant. In addition, the direction of the dispersion tilt
vector also plays an important role in the RKKY effect in titled WSMs
unlike the isotropic RKKY effect in non-tilted WSMs. For both TRS
breaking and IR breaking WSMs, the sign of the RKKY magnetic ordering
can change signs, i.e., from ferromagnetic to anti-ferromagnetic ordering
by varying the azimuthal angle of the displacement vector $\mathrm{\boldsymbol{R}}$
between impurities with respect to the tilt vector, as shown in Fig.
\ref{Fig: trsanalytical}(c) and Fig. \ref{Fig: trsanalytical}(c).
In practical terms, this anisotropy may be utilized in modulating
the sign of the RKKY exchange in layered materials, for example, where
the direction of impurity displacement can be controlled. In bulk
materials where the impurities are distributed isotropically, the
net magnetic ordering is obtained by summing over all the displacement
directions. Thus, one needs to consider the directional dependence
to correctly evaluate the net sign of the RKKY coupling. This anisotropic
dependence of the RKKY coupling with the tilt and separation direction
has not been explored before and can shed light on the magnetic interaction
between impurities and layered structures with WSMs.

\section{Summary}

Using both the analytical method and numerical treatment we study
the effect of tilt in RKKY coupling between magnetic impurities in
Weyl semi-metal for both the TRS breaking and IR-breaking system.
In both cases, the RKKY range function varies to the second order
in the tilt parameter (at the small tilt limit). Additionally, the
tilt-dependent part of the RKKY range function varies as $1/\mathrm{R}$
with the impurity separation distance (in the limit of large $\mathrm{R}$),
compared to the $1/\mathrm{R}^{3}$ dependence of the conventional
RKKY range function in non-tilted WSMs. This means that the tilt-dependent
RKKY term would dominate over the conventional tilt-independent term
beyond a cross-over value of $\mathrm{R}$, typically of the order
of a few nano-meters which is within the practical range of RKKY coupling
in experiments. Finally, the tilt-dependent term in the range function
exhibits a strong anisotropy as the direction of the displacement
vector $\mathrm{R}$ is varied with respect to the tilt direction.
More importantly, the sign of the coupling can change from ferromagnetic
to antiferromagnetic coupling based on the displacement direction.
Our analytical results were verified with numerical calculations.
This work is supported by the Ministry of Education
(MOE) Tier-II grant MOE2018-T2-2-117 (NUS Grant Nos. R-263-000-E45-112/R-398-000-092-112)
and MOE Tier-I FRC grant (NUS Grant No. R-263-000-D66-114).

\bibliographystyle{unsrt}
\bibliography{Bibtex-RKKY-WSM}

\end{document}


\preprint{APS/123-QED}

\title{Supplemental material for ''Ruderman-Kittel-Kasuya-Yosida (RKKY) interaction in Weyl semimetals with tilted energy dispersion''}
\author{Anirban Kundu}
\affiliation{Department of Electrical and Computer Engineering, National University of Singapore, 4 Engineering Drive 3, Singapore 117576}
\affiliation{Physics Department, Ariel University, Ariel 40700, Israel}
\author{M. B. A. Jalil}
\affiliation{Department of Electrical and Computer Engineering, National University of Singapore, 4 Engineering Drive 3, Singapore 117576}


\pacs{}
\maketitle

\section{Method}

RKKY term: The magnetic interaction between conduction electrons and
the impurity can perturbation term can be written as,
\begin{equation}
H'=J\text{ }\sum_{i}\mathbf{S}_{i}\cdot\mathbf{\sigma}\delta(\mathbf{r}-\mathbf{R}_{i})=J\text{ }\left[\mathbf{\sigma}\cdot\mathbf{S}_{1}\delta(\mathbf{r}-\mathbf{R}_{1})+\mathbf{\sigma}\cdot\mathbf{S}_{2}\delta(\mathbf{r}-\mathbf{R}_{2})\right]
\end{equation},
\begin{align}
<\mathbf{k}'\eta|H'|\mathbf{k}\eta'> & =(J/V)\int d\mathbf{r}\text{ }e^{i\left(\mathbf{k}'-\mathbf{Q}\right)\cdot\mathbf{r}}\left[\mathbf{\sigma}\cdot\mathbf{S}_{1}\delta(\mathbf{r}-\mathbf{R}_{1})+\mathbf{\sigma}\cdot\mathbf{S}_{2}\delta(\mathbf{r}-\mathbf{R}_{2})\right]e^{-i\left(\mathbf{k}+\mathbf{Q}\right)\cdot\mathbf{r}}\nonumber \\
 & =(J/V)\left[\mathbf{\sigma}\cdot\mathbf{S}_{1}e^{i(\mathbf{k}'-\mathbf{k}-2\mathbf{Q})\cdot\mathbf{R}_{1}}+\mathbf{\sigma}\cdot\mathbf{S}_{2}e^{i(\mathbf{k'}-\mathbf{k}-2\mathbf{Q})\cdot\mathbf{R}_{2}}\right],
\end{align}
\begin{align}
|<\mathbf{k}\eta|H'|\mathbf{k}'\eta'>|^{2}\left(J/V\right)^{2} & =\left(\mathbf{S}_{1}\right)^{2}+\left(\mathbf{S}_{2}\right)^{2}+2\left(\mathbf{S}_{1}\cdot\mathbf{S}_{2}\right)\cos\left[(\mathbf{k'}-\mathbf{k}-2\mathbf{Q})\cdot\mathbf{R}\right],
\end{align}
 where $\mathbf{q}=\mathbf{k}-\mathbf{k}'$ and $\mathbf{R}=\mathbf{R}_{1}-\mathbf{R}_{2}$.
The energy difference between two bands:

\begin{equation}
\epsilon_{\mathbf{k'},\eta',c'}-\epsilon_{\mathbf{k},\eta,c}=\hbar v_{F}\left(\eta'c'\left|\mathbf{k'}-\eta'\mathbf{Q}\right|-\eta c\left|\mathbf{k}-\eta\mathbf{Q}\right|\right)+\hbar v_{F}\mathbf{\mathbf{w}}_{\eta'}\cdot\left(\mathbf{k}'-\eta'\mathbf{Q}\right)-\hbar v_{F}\mathbf{\mathbf{w}}_{\eta}\cdot\left(\mathbf{k}-\eta\mathbf{Q}\right)+\left(\eta'-\eta\right)\hbar v_{F}Q_{0}.
\end{equation}
Now considering that the Fermi energy $\epsilon_{F}>0$, only the
$\eta=1,c=1$ and $\eta=-1,c=-1$ are relevant. Using the above condition
we obtain, 
\begin{equation}
\epsilon_{\mathbf{k'},\eta'}-\epsilon_{\mathbf{k},\eta}=\hbar v_{F}\left(\left|\mathbf{k'}-\eta'\mathbf{Q}\right|-\left|\mathbf{k}-\eta\mathbf{Q}\right|\right)+\hbar v_{F}\mathbf{\mathbf{w}}_{\eta'}\cdot\left(\mathbf{k}'-\eta'\mathbf{Q}\right)-\hbar v_{F}\mathbf{\mathbf{w}}_{\eta}\cdot\left(\mathbf{k}-\eta\mathbf{Q}\right)+\left(\eta'-\eta\right)\hbar v_{F}Q_{0}.
\end{equation}
The second order correction is $\epsilon^{(2)}=2\left(\mathbf{S}_{1}\cdot\mathbf{S}_{2}\right)F\left(R\right)$,
with,

\begin{equation}
F\left(R\right)=\sum_{\eta}F_{\text{intra}}\left(\eta,R\right)+\sum_{\eta'\neq\eta}e^{i\left(\eta2\mathbf{Q}\cdot\mathbf{R}\right)}F_{\text{inter}}\left(\eta,-\eta,R\right),\label{eq:arb tilt 1}
\end{equation}
where, for intra-valley: $\left(\epsilon_{\mathbf{k'},\eta}-\epsilon_{\mathbf{k},\eta}\right)=\hbar v_{F}\left(\left|\mathbf{k'}-\eta\mathbf{Q}\right|-\left|\mathbf{k}-\eta\mathbf{Q}\right|\right)+\hbar\mathbf{\mathbf{w}}_{\eta}\cdot\left(\mathbf{k}'-\eta\mathbf{Q}\right)-\hbar\mathbf{\mathbf{w}}_{\eta}\cdot\left(\mathbf{k}-\eta\mathbf{Q}\right)$
and we obtain

\begin{align}
F_{\text{intra}}\left(\eta,R\right) & =\text{Re}\left(\frac{1}{\hbar v_{F}}\right)\int_{|\mathbf{k}-\eta\mathbf{Q}|=0}^{k_{F\eta}}\text{d}\mathbf{k}\text{ }\int_{|\mathbf{k}'-\eta\mathbf{Q}|=k_{F\eta}}^{\infty}\text{d}\mathbf{k'}\frac{e^{i\left(\mathbf{k}'-\mathbf{k}\right)\cdot\mathbf{R}}}{\epsilon_{\mathbf{k}'\eta}-\epsilon_{\mathbf{k}\eta}}\nonumber \\
 & =\text{Re}\left(\frac{1}{\hbar v_{F}}\right)\int_{k=0}^{k_{F\eta}}\text{d}\mathbf{k}\text{ }\int_{k'=k_{F\eta}}^{\infty}\text{d}\mathbf{k'}\frac{e^{i\left(\mathbf{k}'-\mathbf{k}\right)\cdot\mathbf{R}}}{\left(k'-k\right)+\frac{\mathbf{\mathbf{w}}_{\eta}}{v_{F}}\cdot\left(\mathbf{k'-k}\right)}.
\end{align}
In the last line above we substituted $\left(\mathbf{k}'-\eta\mathbf{Q}\right)\ensuremath{\rightarrow\mathbf{k}'}$
and $\left(\mathbf{k}-\eta\mathbf{Q}\right)\ensuremath{\rightarrow\mathbf{k}}$without
loss of any generality. For inter-vally ($\eta'\neq\eta$) the energy
difference becomes $\left(\epsilon_{\mathbf{k'},\eta'}-\epsilon_{\mathbf{k},\eta}\right)=\hbar v_{F}\left(\left|\mathbf{k'}-\eta'\mathbf{Q}\right|-\left|\mathbf{k}-\eta\mathbf{Q}\right|\right)+\hbar\mathbf{\mathbf{w}}_{\eta'}\cdot\left(\mathbf{k}'-\eta'\mathbf{Q}\right)-\hbar\mathbf{\mathbf{w}}_{\eta}\cdot\left(\mathbf{k}-\eta\mathbf{Q}\right)+\left(\eta'-\eta\right)\hbar v_{F}q_{0}$
\begin{align}
F_{\text{inter}}\left(\eta,\eta',R\right) & =\text{Re}\int_{0}^{k_{F\eta}}d\mathbf{k}\text{ }\int_{k_{F\eta'}}^{\infty}d\mathbf{k'}\frac{e^{i\left(\mathbf{k}'-\mathbf{k}\right)\cdot\mathbf{R}}}{\epsilon_{\mathbf{k}'\eta'}-\epsilon_{\mathbf{k}\eta}}\\
 & =\text{Re}\int_{\left|\mathbf{k}-\eta\mathbf{Q}\right|=0}^{k_{F\eta}}d\mathbf{k}\text{ }\int_{\left|\mathbf{k'}-\eta'\mathbf{Q}\right|=k_{F\eta'}}^{\infty}d\mathbf{k'}\nonumber \\
 & \times \left[\frac{e^{i\left(\mathbf{k}'-\mathbf{k}\right)\cdot\mathbf{R}}}{\hbar v_{F}\left(\left|\mathbf{k'}-\eta'\mathbf{Q}\right|-\left|\mathbf{k}-\eta\mathbf{Q}\right|\right)+\hbar\mathbf{\mathbf{w}}_{\eta'}\cdot\left(\mathbf{k}'-\eta'\mathbf{Q}\right)-\hbar\mathbf{\mathbf{w}}_{\eta}\cdot\left(\mathbf{k}-\eta\mathbf{Q}\right)+\left(\eta'-\eta\right)\hbar v_{F}q_{0}}\right]\\
 & =\text{Re}\int_{k=0}^{k_{F\eta}}d\mathbf{k}\text{ }\int_{k'=k_{F\eta'}}^{\infty}d\mathbf{k'}\frac{e^{i\left(\mathbf{k}'-\mathbf{k}\right)\cdot\mathbf{R}}}{\hbar v_{F}\left(k'-k\right)+\hbar\left(\mathbf{\mathbf{w}}_{\eta'}\cdot\mathbf{k}'-\hbar v_{F}\mathbf{\mathbf{w}}_{\eta}\cdot\mathbf{k}\right)+\left(\eta'-\eta\right)\hbar v_{F}q_{0}}\label{eq:arb tilt 2}.
\end{align}
Finally we can write using $\eta'=-\eta$,
\begin{equation}
F_{\text{inter}}\left(\eta,-\eta,R\right)=\text{Re}\int_{k=0}^{k_{F\eta}}d\mathbf{k}\text{ }\int_{k'=k_{F-\eta}}^{\infty}d\mathbf{k'}\frac{e^{i\left(\mathbf{k}'-\mathbf{k}\right)\cdot\mathbf{R}}}{\hbar v_{F}\left(k'-k\right)+\hbar\left(\mathbf{\mathbf{w}}_{\mp}\cdot\mathbf{k}'-\hbar v_{F}\mathbf{\mathbf{w}}_{\pm}\cdot\mathbf{k}\right)-\eta2\hbar v_{F}q_{0}}\label{eq:arb tilt 3}.
\end{equation}
\section{Small tilt limit }
 As stated before,
\begin{equation}
F\left(R\right)=\sum_{\eta}F_{\text{intra}}\left(\eta,R\right)+\sum_{\eta'\neq\eta}e^{i\left(\eta2\mathbf{Q}\cdot\mathbf{R}\right)}F_{\text{inter}}\left(\eta,-\eta,R\right),
\end{equation}
Expanding to the second order in tilt:
\begin{align}
F_{\text{intra}}\left(\eta,R\right) & =\text{Re}\int_{k=0}^{k_{F\eta}}\text{d}\mathbf{k}\text{ }\int_{k'=k_{F\eta}}^{\infty}\text{d}\mathbf{k'}\frac{e^{i\left(\mathbf{k}'-\mathbf{k}\right)\cdot\mathbf{R}}}{\hbar v_{F}\left(k'-k\right)}+\text{Re}\int_{k=0}^{k_{F\eta}}\text{d}\mathbf{k}\text{ }\int_{k'=k_{F\eta}}^{\infty}\text{d}\mathbf{k'}\frac{\left(\mathbf{w}_{\eta}\cdot\mathbf{k}'-\mathbf{w}_{-\eta}\cdot\mathbf{k}\right)^{2}e^{i\left(\mathbf{k}'-\mathbf{k}\right)\cdot\mathbf{R}}}{\hbar v_{F}\left(k'-k\right)^{3}}\\
 & =\sum_{m=0}^{2}\left(\frac{\mathbf{w}_{\pm}}{v_{F}}\cdot\frac{\partial}{\partial\mathbf{R}}\right)^{m}\chi^{(m)}\left(\mathrm{R}\right),
\end{align}
where, 
\begin{equation}
\chi^{(m)}\left(\mathrm{R}\right)=\left(\frac{1}{\hbar v_{F}}\right)\text{Re}\left(-i\right)^{m}\int_{0}^{k_{F\eta}}\text{d}\mathbf{k}\text{ }\int_{k_{F\eta}}^{\infty}\text{d}\mathbf{k}'\frac{e^{i\left(\mathbf{k}'-\mathbf{k}\right)\cdot\mathbf{R}}}{\left(k'-k\right)^{m+1}}.\label{eq:Fintramainchi}
\end{equation}
And the inter-valley range function becomes,

\begin{align}
F_{\text{inter}}^{\eta,-\eta}\left(\mathrm{R}\right) & =\sum_{m=0}^{2}\zeta^{(m,\eta,-\eta)}\left(\mathrm{R}\right),\label{eq:Fintermain}
\end{align}
where, 
\begin{equation}
\zeta^{(m,\eta,-\eta)}\left(\mathrm{R}\right)=\text{Re}\int_{k=0}^{k_{F\eta}}d\mathbf{k}\text{ }\int_{k'=k_{F,-\eta}}^{\infty}d\mathbf{k}'\frac{\hbar^{m}\left(\mathbf{w}_{\eta}\cdot\mathbf{k}'-\mathbf{w}_{-\eta}\cdot\mathbf{k}\right)^{m}e^{i\left(\mathbf{k}'-\mathbf{k}\right)\cdot\mathbf{R}}}{\left(\hbar v_{F}\left(k'-k\right)\mp2\epsilon_{0}\right)^{m+1}}.\label{eq:Fintermain-1}
\end{equation}
Clearly for chiral tilt i.e., $\mathbf{w}_{\pm}=\pm\mathbf{w}$
the second term vanishes after summing over the contribution for a
pair of nodes. Also the imaginary part of the integration Im$[...]$
is zero, meaning that for $\mathbf{w}_{\pm}=\mathbf{w}$ the
linear in tilt term vanishes. Finally we obtain in general for intra-valley,
putting $q_{0}=0$ in the above case leads to the coupling for TRS
symmetry breaking case, however $q_{0}\neq0$ implies IR symmetry
breaking case. Results for parallel tilt\textit{ ($\mathbf{w}_{\eta}=\mathbf{w}$),}

\subsection{TRS broken ($\ensuremath{\epsilon_{0}=0} \text{ and }  \ensuremath{\mathbf{Q}\protect\ne0}$)}

The Hamiltonian in that scenario is, $H_{0}=\eta v_{F}\mathbf{\tau}\cdot\left(\mathbf{k}-\eta\mathbf{Q}\right)+\eta\mathbf{w}\cdot\left(\mathbf{k}-\eta\mathbf{Q}\right)\tau_{0}$
and $k_{F\eta}=k_{F}=\epsilon_{F}/\hbar v_{F}$ we get

\begin{align}
F_{\text{intra}}\left(\eta,R\right) & =\text{Re}\int_{k=0}^{k_{F\eta}}\text{d}\mathbf{k}\text{ }\int_{k'=k_{F\eta}}^{\infty}\text{d}\mathbf{k'}\frac{e^{i\left(\mathbf{k}'-\mathbf{k}\right)\cdot\mathbf{R}}}{\hbar v_{F}\left(k'-k\right)}\nonumber \\
 & +\hbar\left(\mathbf{w}\cdot\mathbf{\nabla}_{\mathbf{R}}\right)^{2}\text{Re}\int_{k=0}^{k_{F\eta}}\text{d}\mathbf{k}\text{ }\int_{k'=k_{F\eta}}^{\infty}\text{d}\mathbf{k'}\frac{e^{i\left(\mathbf{k}'-\mathbf{k}\right)\cdot\mathbf{R}}}{\left[\hbar v_{F}\left(k'-k\right)\right]^{3}}\\
 & =\sum_{m=0}^{2}\left(\frac{\mathbf{w}}{v_{F}}\cdot\frac{\partial}{\partial\mathbf{R}}\right)^{m}\chi^{(m,\eta)}\left(R\right)
,\end{align}
where,
\begin{align}
\chi^{(m,\eta)}\left(R\right)=\left(\frac{1}{\hbar v_{F}}\right)\text{Re}\left(-i\right)^{m}\int_{0}^{k_{F}}\text{d}\mathbf{k}\text{ }\int_{k_{F}}^{\infty}\text{d}\mathbf{k'}\frac{e^{i\left(\mathbf{k}'-\mathbf{k}\right)\cdot\mathbf{R}}}{\left(k'-k\right)^{m+1}},
\end{align}

\textit{
\begin{align}
\zeta^{(m)}\left(\eta,-\eta,R\right) & =\text{Re}\int_{k=0}^{k_{F\eta}}d\mathbf{k}\text{ }\int_{k'=k_{F-\eta}}^{\infty}d\mathbf{k'}\frac{\hbar^{m}\left(\mathbf{w}\cdot\mathbf{k}'-\mathbf{w}\cdot\mathbf{k}\right)^{m}e^{i\left(\mathbf{k}'-\mathbf{k}\right)\cdot\mathbf{R}}}{\left[\hbar v_{F}\left(k'-k\right)\right]^{m+1}}\\
 & =\left(\frac{1}{\hbar v_{F}}\right)\left(\frac{\mathbf{w}}{v_{F}}\cdot\frac{\partial}{\partial\mathbf{R}}\right)^{m}\text{Re}\left(-i\right)^{m}\int_{k=0}^{k_{F}}d\mathbf{k}\text{ }\int_{k'=k_{F}}^{\infty}d\mathbf{k'}\frac{e^{i\left(\mathbf{k}'-\mathbf{k}\right)\cdot\mathbf{R}}}{\left(k'-k\right)^{m+1}}\\
 & =\left(\frac{\mathbf{w}}{v_{F}}\cdot\frac{\partial}{\partial\mathbf{R}}\right)^{m}\chi^{(m,\eta)}\left(R\right)
\end{align}
}The inter contribution is,

\begin{equation}
F_{\text{inter}}^{\eta,-\eta}\left(\mathrm{R}\right)=\left(\frac{\mathbf{w}}{v_{F}}\cdot\frac{\partial}{\partial\mathbf{R}}\right)^{m}\chi^{(m,\eta)}\left(R\right)
\end{equation}
So, the total range function is, 
\begin{align}
F\left(\mathbf{R}\right) & =\sum_{\eta}F_{\text{intra}}^{\eta}\left(\mathrm{R}\right)+\sum_{\eta}e^{i\left(\eta2\mathbf{Q}\cdot\mathbf{R}\right)}F_{\text{inter}}^{\eta,-\eta}\left(\mathrm{R}\right).\\
 & =2\left(1+\cos\left(2\mathbf{Q}\cdot\mathbf{R}\right)\right)\left(\chi^{(0,\eta)}\left(\mathrm{R}\right)-\left(\frac{\mathbf{w}}{v_{F}}\cdot\frac{\partial}{\partial\mathbf{R}}\right)^{2}\chi^{(2,\eta)}\left(\mathrm{R}\right)\right)
\end{align}
\subsection{IR-broken ($\ensuremath{\mathbf{Q}=0}\text{ and }\ensuremath{\epsilon\protect\neq0}$)}
The Hamiltonian is: $H_{0}=\eta v_{F}\mathbf{\tau}\cdot\mathbf{k}+\mathbf{w}_{\eta}\cdot\mathbf{k}\tau_{0}+\eta Q_{0}\tau_{0}$.
The total range function,
\begin{align}
F^{IR}\left(\mathbf{R}\right) & =\left\{ \sum_{\eta}\chi_{IR}^{(0,\eta)}\left(\mathrm{R}\right)-\sum_{\eta}\left(\frac{\mathbf{w}}{v_{F}}\cdot\frac{\partial}{\partial\mathbf{R}}\right)^{2}\chi_{IR}^{(2,\eta)}\left(\mathrm{R}\right)\right\} \nonumber \\
 & +\left\{ \sum_{\eta}\left[\zeta_{IR}^{(0)}\left(\eta,-\eta,\mathrm{R}\right)-\left(\frac{\mathbf{w}}{v_{F}}\cdot\frac{\partial}{\partial\mathbf{R}}\right)^{2}\zeta_{IR}^{(2)}\left(\eta,-\eta,\mathrm{R}\right)\right]\right\} ,
\end{align}
and we introduce,
\begin{align}
\chi^{(n,\pm)}\left(R\right) & =\left(\frac{1}{\hbar v_{F}}\right)\text{Re}\left[\left(-i\right)^{n}\int_{0}^{k_{F\pm}}\text{d}\mathbf{k}\text{ }\int_{k_{F\pm}}^{\infty}\text{d}\mathbf{k'}\frac{e^{i\left(\mathbf{k}'-\mathbf{k}\right)\cdot\mathbf{R}}}{\left(k'-k\right)^{n+1}}\right],
\end{align}
and 
\begin{align}
\zeta_{IR}^{(n)}\left(\eta,-\eta,\mathrm{R}\right) & =\text{Re}\int_{k=0}^{k_{F\eta}}d\mathbf{k}\text{ }\int_{k'=k_{F\eta'}}^{\infty}d\mathbf{k'}\frac{e^{i\left(\mathbf{k}'-\mathbf{k}\right)\cdot\mathbf{R}}}{\left(k'-k\right)\mp2q_{0}^{n+1}}\\
 & \text{using, }\int d\mathbf{k}=\int\text{d}k\text{ }k^{2}\int\text{d\ensuremath{\theta}}\sin\theta\int\text{d}\phi=\int\text{d}k\text{ }k^{2}\int_{-1}^{1}\text{d}x\int\text{d}\phi=\nonumber \\
 & =\left(\frac{4\pi}{R}\right)^{2}\int_{0}^{k_{F\eta}}dk\text{ }\left(k\sin kR\right)\text{ }\int_{k_{F\eta}}^{\infty}dk'\frac{\text{ }k'\sin k'R}{\left(k'-k\pm2q_{0}\right)}\nonumber \\
 & \text{Rescale }k\rightarrow k-\eta q_{0}\text{ such that }\int_{0}^{k_{F\pm}}dk\rightarrow\int_{0}^{k_{F}}dk\text{ cause }k_{F}=k_{F\pm}+\eta q_{0}\nonumber \\
 & =\left(\frac{4\pi}{R}\right)^{2}\int_{0}^{k_{F}}dk\text{ }\left(k-\eta q_{0}\right)\sin\left[\left(k-\eta q_{0}\right)R\right]\text{ }\int_{k_{F}}^{\infty}dk'\frac{\left(k'\pm q_{0}\right)\sin\left[\left(k'+\eta q_{0}\right)R\right]}{\left(k'-k\right)^{3}}\\
 & =\left(\frac{4\pi}{R}\right)^{2}\int_{0}^{k_{F}}dk\text{ }\left(k-\eta q_{0}\right)\sin\left[\left(k-\eta q_{0}\right)R\right]\text{ }\int_{0}^{\infty}dk'\frac{\left(k'\pm q_{0}\right)\sin\left[\left(k'+\eta q_{0}\right)R\right]}{\left(k'-k\right)^{3}}.
\end{align}
After contour integration,
\begin{align}
\sum_{\eta}\chi_{IR}^{(0,\eta)}\left(\mathrm{R}\right) & =\frac{2\pi^{2}\left(\left(1-2R^{2}(\text{\ensuremath{k_{F}}}-\text{\ensuremath{q_{0}}})^{2}\right)\cos(2R(\text{\ensuremath{k_{F}}}-\text{\ensuremath{q_{0}}}))+\left(1-2R^{2}(\text{\ensuremath{k_{F}}}+\text{\ensuremath{q_{0}}})^{2}\right)\cos(2R(\text{\ensuremath{k_{F}}}+\text{\ensuremath{q_{0}}}))\right)}{R^{5}}\nonumber \\
 & +\frac{2\pi^{2}\left(+2R(\text{\ensuremath{k_{F}}}-\text{\ensuremath{q_{0}}})\sin(2R(\text{\ensuremath{k_{F}}}-\text{\ensuremath{q_{0}}}))+2R(\text{\ensuremath{k_{F}}}+\text{\ensuremath{q_{0}}})\sin(2R(\text{\ensuremath{k_{F}}}+\text{\ensuremath{q_{0}}}))-2\right)}{R^{5}},
\end{align}
\begin{align}
\sum_{\eta}\zeta^{(0,\eta,-\eta)}\left(R\right) & =\frac{\pi\left(\left(1-2R^{2}(\text{\ensuremath{k_{F}}}-\text{\ensuremath{q_{0}}})(\text{\ensuremath{k_{F}}}+\text{\ensuremath{q_{0}}})\right)\cos(2\text{\ensuremath{k_{F}}}R)+2\text{\ensuremath{k_{F}}}R\sin(2\text{\ensuremath{k_{F}}}R)-2\text{\ensuremath{q_{0}}}^{2}R^{2}-1\right)}{2R^{3}},
\end{align}
\begin{align}
 & \sum_{\eta}\chi_{IR}^{(0,\eta)}\left(\mathrm{R}\right)+\sum_{\eta}\zeta^{(0,\eta,-\eta)}\left(R\right)\nonumber \\
 & =\frac{2\pi^{2}\left(2\left(1-2R^{2}(\text{\ensuremath{k_{F}}}-\text{\ensuremath{q_{0}}})(\text{\ensuremath{k_{F}}}+\text{\ensuremath{q_{0}}})\right)\cos(2\text{\ensuremath{k_{F}}}R)+\left(1-2R^{2}(\text{\ensuremath{k_{F}}}-\text{\ensuremath{q_{0}}})^{2}\right)\cos(2R(\text{\ensuremath{k_{F}}}-\text{\ensuremath{q_{0}}}))\right)}{R^{5}}\nonumber \\
 & +\frac{2\pi^{2}\left(\left(1-2R^{2}(\text{\ensuremath{k_{F}}}+\text{\ensuremath{q_{0}}})^{2}\right)\cos(2R(\text{\ensuremath{k_{F}}}+\text{\ensuremath{q_{0}}}))+2R(\text{\ensuremath{k_{F}}}-\text{\ensuremath{q_{0}}})\sin(2R(\text{\ensuremath{k_{F}}}-\text{\ensuremath{q_{0}}}))\right)}{R^{5}}\nonumber \\
 & +\frac{2\pi^{2}\left(2R(\text{\ensuremath{k_{F}}}+\text{\ensuremath{q_{0}}})\sin(2R(\text{\ensuremath{k_{F}}}+\text{\ensuremath{q_{0}}}))+4\text{\ensuremath{k_{F}}}R\sin(2\text{\ensuremath{k_{F}}}R)-4\text{\ensuremath{q_{0}}}^{2}R^{2}-4\right)}{R^{5}}.
\end{align}
Contribution from the tilt,
\begin{align}
\chi_{IR}^{(2,\eta)}\left(\mathrm{R}\right) & =\frac{4\pi^{3}\left(-4R^{2}(q_{F}-\eta q_{0})^{2}+\left(2R^{2}(q_{F}-\eta q_{0})^{2}+1\right)\cos(2R(q_{F}-\eta q_{0}))+2R(q_{F}-\eta q_{0})\sin(2R(q_{F}-\eta q_{0}))-1\right)}{R^{3}},
\end{align}
\begin{align}
\zeta_{IR}^{(2)}\left(\eta,-\eta,\mathrm{R}\right) & =\frac{4\pi^{3}\left(2R\left(3\eta q_{0}-2R^{2}(k_{F}-4\eta q_{0})(k_{F}-\eta q_{0})^{2}\right)\sin(2\eta q_{0}R)-3\left(4R^{2}(k_{F}-\eta q_{0})^{2}+1\right)\cos(2\eta q_{0}R)\right)}{3R^{3}}\nonumber\\
 & +\frac{4\pi^{3}\left(3\left(2R^{2}(k_{F}-\eta q_{0})(k_{F}-3\eta q_{0})+1\right)\cos(2R(k_{F}-2\eta q_{0}))+6k_{F}R\sin(2R(k_{F}-2\eta q_{0}))\right)}{3R^{3}}.
\end{align}
For $\mathbf{w}\parallel\hat{z}$
\begin{align}
 & \left(\frac{|\mathbf{w}|}{v_{F}}\cdot\frac{\partial}{\partial R_{z}}\right)^{2}\chi_{IR}^{(2,\eta)}\left(\mathrm{R}\right)\nonumber\\
 & =-\frac{2\pi^{3}\left(R^{2}\left(-4\left(R^{2}-3R_{z}^{2}\right)(k_{\text{F}}-\eta q_0 )^{2}-3\right)+15R_{z}^{2}\right)}{R^{7}}\nonumber\\
 & -\frac{2\pi^{3}\left(+2R( k_{\text{F}} -\eta q_0 )\sin(2R( k_{\text{F}} -\eta q_0 ))\left(R^{2}\left(2(R-R_{z})(R+R_{z})( k_{\text{F}} -\eta q_0 )^{2}+3\right)-15R_{z}^{2}\right)\right)}{R^{7}}\nonumber\\
 & -\frac{2\pi^{3}\left(\cos(2R( k_{\text{F}} -\eta q_0 ))\left(R^{2}\left(2( k_{\text{F}} -\eta q_0 )^{2}\left(R^{2}(2 k_{\text{F}} R_{z}-2\eta q_0 R_{z}-1)(2 k_{\text{F}} R_{z}-2\eta q_0 R_{z}+1)+9R_{z}^{2}\right)+3\right)-15R_{z}^{2}\right)\right)}{R^{7}},
\end{align}
and 
\begin{small}
\begin{align}
 & \left(\frac{|\mathbf{w}|}{v_{F}}\cdot\frac{\partial}{\partial R_{z}}\right)^{2}\zeta_{IR}^{(2)}\left(\eta,-\eta,\mathrm{R}\right)\nonumber\\
 & =\frac{2\pi^{3}\left(-3\cos(2 k_{\text{F}} R)\left(R^{2}\left(2z^{2}\left(9 k_{\text{F}} ^{2}-20\eta k_{\text{F}}  q_0 +3\eta^{2} q_0 ^{2}\right)+3\right)+R^{4}\left(8 k_{\text{F}} ^{4}z^{2}-2\eta^{2} q_0 ^{2}\left(4 k_{\text{F}} ^{2}z^{2}+1\right)-2 k_{\text{F}} ^{2}+8\eta k_{\text{F}}  q_0 \right)-15z^{2}\right)\right)}{3R^{7}}\nonumber\\
 & +\frac{2\pi^{3}\left(3R^{2}\left(3-2\eta^{2} q_0 ^{2}\left(R^{2}-3z^{2}\right)\right)-45z^{2}\right)}{3R^{7}}\nonumber\\
 & +\frac{2\pi^{3}\left(4 k_{\text{F}} R^{2}\cos(2\eta q_0 R)\left(2\eta q_0 \left(R^{2}\left( k_{\text{F}} ^{2}(R-z)(R+z)-3\right)+9z^{2}\right)+12\eta^{2} k_{\text{F}}  q_0 ^{2}R^{2}z^{2}+3 k_{\text{F}} \left(R^{2}-3z^{2}\right)-6\eta^{3} q_0 ^{3}R^{2}\left(R^{2}+3z^{2}\right)\right)\right)}{3R^{7}}\nonumber\\
 & \frac{2\pi^{3}\left(-8\eta k_{\text{F}}  q_0 R^{3}\sin(2\eta q_0 R)\left(2\eta q_0 \left(R^{2}\left( k_{\text{F}} ^{2}z^{2}+3\right)-9z^{2}\right)-3 k_{\text{F}} \left(R^{2}-3z^{2}\right)-6\eta^{3} q_0 ^{3}R^{2}z^{2}\right)\right)}{3R^{7}}\nonumber\\
 & \frac{2\pi^{3}\left(-6R\sin(2 k_{\text{F}} R)\left(2 k_{\text{F}} ^{3}R^{2}(R-z)(R+z)-4\eta q_0 \left(2z^{2}\left( k_{\text{F}} ^{2}R^{2}-2\right)+R^{2}\right)-2\eta^{2} k_{\text{F}}  q_0 ^{2}R^{2}\left(R^{2}-3z^{2}\right)+3 k_{\text{F}} \left(R^{2}-5z^{2}\right)\right)\right)}{3R^{7}}.
\end{align}
\end{small}
Finally the full expression for range function for IR broken system is as follows,
\begin{small}
\begin{align}
F^{IR}\left(\mathbf{R}\right) 
 & =\frac{2\pi^{2}\left(2\left(1-2R^{2}(\text{\ensuremath{k_{F}}}-\text{\ensuremath{q_{0}}})(\text{\ensuremath{k_{F}}}+\text{\ensuremath{q_{0}}})\right)\cos(2\text{\ensuremath{k_{F}}}R)+\left(1-2R^{2}(\text{\ensuremath{k_{F}}}-\text{\ensuremath{q_{0}}})^{2}\right)\cos(2R(\text{\ensuremath{k_{F}}}-\text{\ensuremath{q_{0}}}))\right)}{R^{5}}\nonumber \\
 & +\frac{2\pi^{2}\left(\left(1-2R^{2}(\text{\ensuremath{k_{F}}}+\text{\ensuremath{q_{0}}})^{2}\right)\cos(2R(\text{\ensuremath{k_{F}}}+\text{\ensuremath{q_{0}}}))+2R(\text{\ensuremath{k_{F}}}-\text{\ensuremath{q_{0}}})\sin(2R(\text{\ensuremath{k_{F}}}-\text{\ensuremath{q_{0}}}))\right)}{R^{5}}\nonumber \\
 & +\frac{2\pi^{2}\left(2R(\text{\ensuremath{k_{F}}}+\text{\ensuremath{q_{0}}})\sin(2R(\text{\ensuremath{k_{F}}}+\text{\ensuremath{q_{0}}}))+4\text{\ensuremath{k_{F}}}R\sin(2\text{\ensuremath{k_{F}}}R)-4\text{\ensuremath{q_{0}}}^{2}R^{2}-4\right)}{R^{5}}\nonumber\\
 &+ \frac{|\mathbf{w}|^2}{v^2_{F}}[...]\nonumber\\
 &\text{with}\nonumber\\
& [...]\nonumber\\
 & =-\frac{2\pi^{3}\left(R^{2}\left(-4\left(R^{2}-3R_{z}^{2}\right)(k_{\text{F}}-\eta q_0 )^{2}-3\right)+15R_{z}^{2}\right)}{R^{7}}\nonumber\\
 & -\frac{2\pi^{3}\left(+2R( k_{\text{F}} -\eta q_0 )\sin(2R( k_{\text{F}} -\eta q_0 ))\left(R^{2}\left(2(R-R_{z})(R+R_{z})( k_{\text{F}} -\eta q_0 )^{2}+3\right)-15R_{z}^{2}\right)\right)}{R^{7}}\nonumber\\
 & -\frac{2\pi^{3}\left(\cos(2R( k_{\text{F}} -\eta q_0 ))\left(R^{2}\left(2( k_{\text{F}} -\eta q_0 )^{2}\left(R^{2}(2 k_{\text{F}} R_{z}-2\eta q_0 R_{z}-1)(2 k_{\text{F}} R_{z}-2\eta q_0 R_{z}+1)+9R_{z}^{2}\right)+3\right)-15R_{z}^{2}\right)\right)}{R^{7}}\nonumber\\
\nonumber \\
  & +\frac{2\pi^{3}\left(-3\cos(2 k_{\text{F}} R)\left(R^{2}\left(2z^{2}\left(9 k_{\text{F}} ^{2}-20\eta k_{\text{F}}  q_0 +3\eta^{2} q_0 ^{2}\right)+3\right)+R^{4}\left(8 k_{\text{F}} ^{4}z^{2}-2\eta^{2} q_0 ^{2}\left(4 k_{\text{F}} ^{2}z^{2}+1\right)-2 k_{\text{F}} ^{2}+8\eta k_{\text{F}}  q_0 \right)-15z^{2}\right)\right)}{3R^{7}}\nonumber\\
 & +\frac{2\pi^{3}\left(3R^{2}\left(3-2\eta^{2} q_0 ^{2}\left(R^{2}-3z^{2}\right)\right)-45z^{2}\right)}{3R^{7}}\nonumber\\
 & +\frac{2\pi^{3}\left(4 k_{\text{F}} R^{2}\cos(2\eta q_0 R)\left(2\eta q_0 \left(R^{2}\left( k_{\text{F}} ^{2}(R-z)(R+z)-3\right)+9z^{2}\right)+12\eta^{2} k_{\text{F}}  q_0 ^{2}R^{2}z^{2}+3 k_{\text{F}} \left(R^{2}-3z^{2}\right)-6\eta^{3} q_0 ^{3}R^{2}\left(R^{2}+3z^{2}\right)\right)\right)}{3R^{7}}\nonumber\\
 & \frac{2\pi^{3}\left(-8\eta k_{\text{F}}  q_0 R^{3}\sin(2\eta q_0 R)\left(2\eta q_0 \left(R^{2}\left( k_{\text{F}} ^{2}z^{2}+3\right)-9z^{2}\right)-3 k_{\text{F}} \left(R^{2}-3z^{2}\right)-6\eta^{3} q_0 ^{3}R^{2}z^{2}\right)\right)}{3R^{7}}\nonumber\\
 & \frac{2\pi^{3}\left(-6R\sin(2 k_{\text{F}} R)\left(2 k_{\text{F}} ^{3}R^{2}(R-z)(R+z)-4\eta q_0 \left(2z^{2}\left( k_{\text{F}} ^{2}R^{2}-2\right)+R^{2}\right)-2\eta^{2} k_{\text{F}}  q_0 ^{2}R^{2}\left(R^{2}-3z^{2}\right)+3 k_{\text{F}} \left(R^{2}-5z^{2}\right)\right)\right)}{3R^{7}}.
\end{align}
\end{small}

\section{Oppositely tilted (\textit{$\mathbf{w}_{\eta}=\pm\mathbf{w}$)}}

\textbf{TRS broken ($Q_{0}=0$):}\textit{ }
\begin{align}
\chi^{(2,\eta)}\left(\mathrm{R}\right) & =-\text{Re}\int_{k=0}^{k_{F}}d\mathbf{k}\text{ }\int_{k'=0}^{\infty}d\mathbf{k'}\frac{\left(\mathbf{w}\cdot\left(\mathbf{k'}+\mathbf{k}\right)\right)^{2}e^{i\left(\mathbf{k}'-\mathbf{k}\right)\cdot\mathbf{R}}}{\left(\hbar v_{F}\left(k'-k\right)\right)^{3}}\\
 & =\text{Re}\left(\int_{k=0}^{k_{F}}d\mathbf{k}\text{ }\left(\frac{\mathbf{w}}{v_{F}}\cdot\frac{\partial}{\partial\mathbf{R}}\right)^{2}e^{-i\mathbf{k}\cdot\mathbf{R}}\right)\int_{k'=0}^{\infty}d\mathbf{k'}\frac{e^{i\mathbf{k}'\cdot\mathbf{R}}}{\left(\hbar v_{F}\left(k'-k\right)\right)^{3}}\nonumber\\
 & +\text{Re}\int_{k=0}^{k_{F}}d\mathbf{k}\text{ }e^{-i\mathbf{k}\cdot\mathbf{R}}\left[\left(\frac{\mathbf{w}}{v_{F}}\cdot\frac{\partial}{\partial\mathbf{R}}\right)^{2}\int_{k'=0}^{\infty}d\mathbf{k'}\frac{e^{i\mathbf{k}'\cdot\mathbf{R}}}{\left(\hbar v_{F}\left(k'-k\right)\right)^{3}}\right]\nonumber\\
 & +2\text{Re}\left[\left(\frac{\mathbf{w}}{v_{F}}\cdot\frac{\partial}{\partial\mathbf{R}}\right)\int_{k=0}^{k_{F}}d\mathbf{k}\text{ }e^{-i\mathbf{k}\cdot\mathbf{R}}\right]\left[\left(\frac{\mathbf{w}}{v_{F}}\cdot\frac{\partial}{\partial\mathbf{R}}\right)\int_{k'=0}^{\infty}d\mathbf{k'}\frac{e^{i\mathbf{k}'\cdot\mathbf{R}}}{\left(\hbar v_{F}\left(k'-k\right)\right)^{3}}\right].
\end{align}
We calculate above but as the final expression is too large, we just
show plot the results in manuscript.
\textbf{IR-broken ($\mathbf{Q}=0$):} 
\begin{align}
F\left(\mathbf{R}\right) & =\left\{ \sum_{\eta}\chi^{(0,\eta)}\left(\mathrm{R}\right)-\sum_{\eta}\chi^{(2,\eta)}\left(\mathrm{R}\right)\right\}\nonumber \\
 & +\left\{ \sum_{\eta}\left[\xi^{(0,\eta,-\eta)}\left(q_{0},\mathrm{R}\right)-\xi^{(2,\eta,-\eta)}\left(q_{0},\mathrm{R}\right)\right]\right\},
\end{align}
with
\textbf{\textit{
\begin{align}
\zeta^{(2)}\left(\eta,-\eta,\mathrm{R}\right) & =-\text{Re}\int_{k=0}^{k_{F\eta}}d\mathbf{k}\text{ }\int_{k'=0}^{\infty}d\mathbf{k'}\frac{\left(\mathbf{w}\cdot\left(\mathbf{k'}+\mathbf{k}\right)\right)^{2}e^{i\left(\mathbf{k}'-\mathbf{k}\right)\cdot\mathbf{R}}}{\left(\hbar v_{F}\left(k'-k\right)\mp2q_{0}\right)^{3}}\\
 & =\text{Re}\left(\left(\frac{\mathbf{w}}{v_{F}}\cdot\frac{\partial}{\partial\mathbf{R}}\right)^{2}\int_{k=0}^{k_{F\eta}}d\mathbf{k}\text{ }e^{-i\mathbf{k}\cdot\mathbf{R}}\right)\int_{k'=0}^{\infty}d\mathbf{k'}\frac{e^{i\mathbf{k}'\cdot\mathbf{R}}}{\left(\hbar v_{F}\left(k'-k\right)\mp2q_{0}\right)^{3}}\nonumber\\
 & +\text{Re}\int_{k=0}^{k_{F\eta}}d\mathbf{k}\text{ }e^{-i\mathbf{k}\cdot\mathbf{R}}\left[\left(\frac{\mathbf{w}}{v_{F}}\cdot\frac{\partial}{\partial\mathbf{R}}\right)^{2}\int_{k'=0}^{\infty}d\mathbf{k'}\frac{e^{i\mathbf{k}'\cdot\mathbf{R}}}{\left(\hbar v_{F}\left(k'-k\right)\mp2q_{0}\right)^{3}}\right]\nonumber\\
 & +2\text{Re}\left[\left(\frac{\mathbf{w}}{v_{F}}\cdot\frac{\partial}{\partial\mathbf{R}}\right)\int_{k=0}^{k_{F\eta}}d\mathbf{k}\text{ }e^{-i\mathbf{k}\cdot\mathbf{R}}\right]\left[\left(\frac{\mathbf{w}}{v_{F}}\cdot\frac{\partial}{\partial\mathbf{R}}\right)\int_{k'=0}^{\infty}d\mathbf{k'}\frac{e^{i\mathbf{k}'\cdot\mathbf{R}}}{\left(\hbar v_{F}\left(k'-k\right)\mp2q_{0}\right)^{3}}\right],
\end{align}
}}
and, 
\begin{align}
\chi^{(2,\eta)}\left(\mathrm{R}\right)=\left.\zeta^{(2)}\left(\eta,\eta',R\right)\right|_{q_{0}=0}.
\end{align}
For the larger size of the expressions we do not show the expressions
here.
\begin{figure}
\begin{centering}
\subfloat[]{\includegraphics[width=0.5\linewidth]{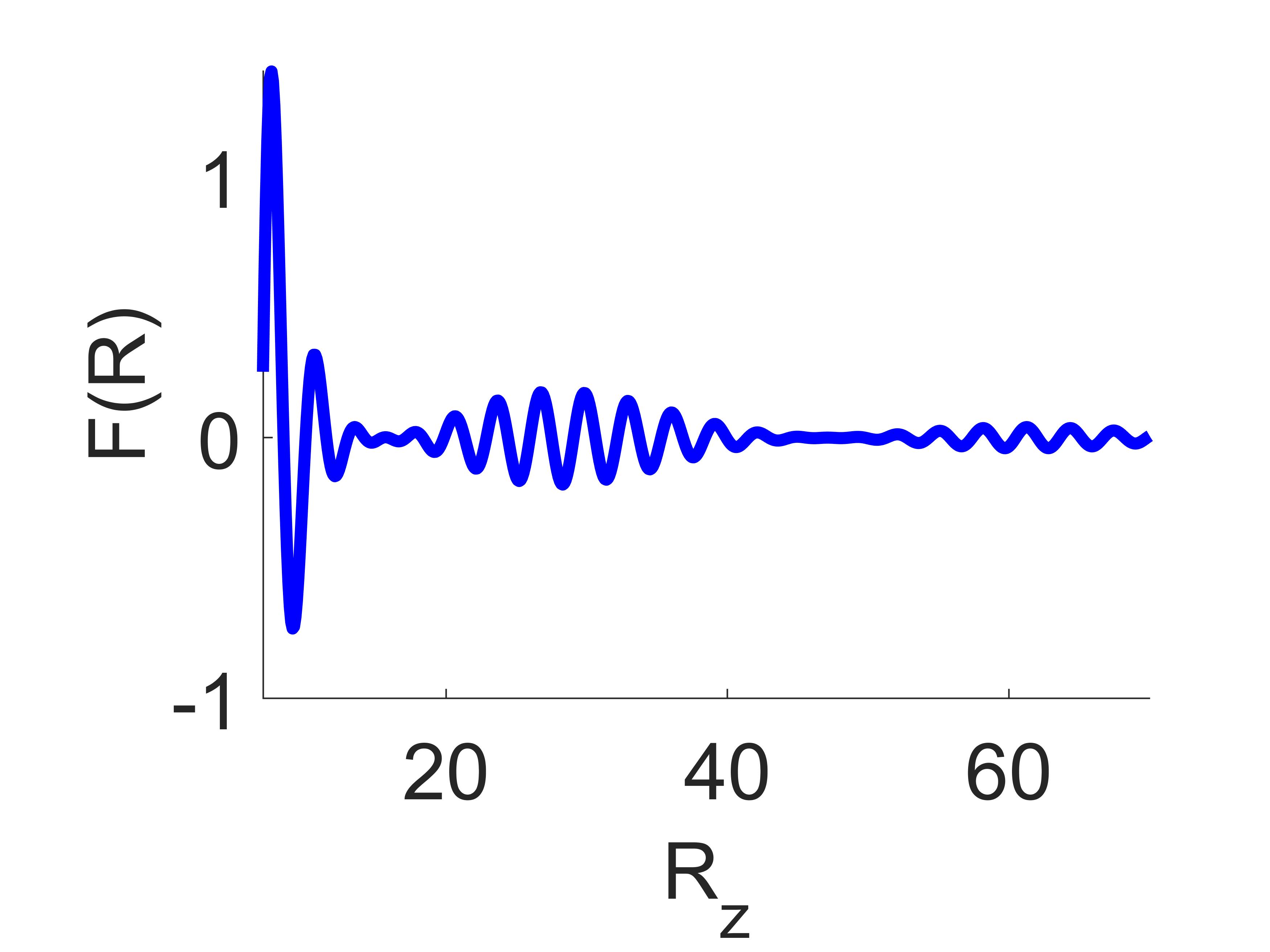}}
\subfloat[]{\includegraphics[width=0.5\linewidth]{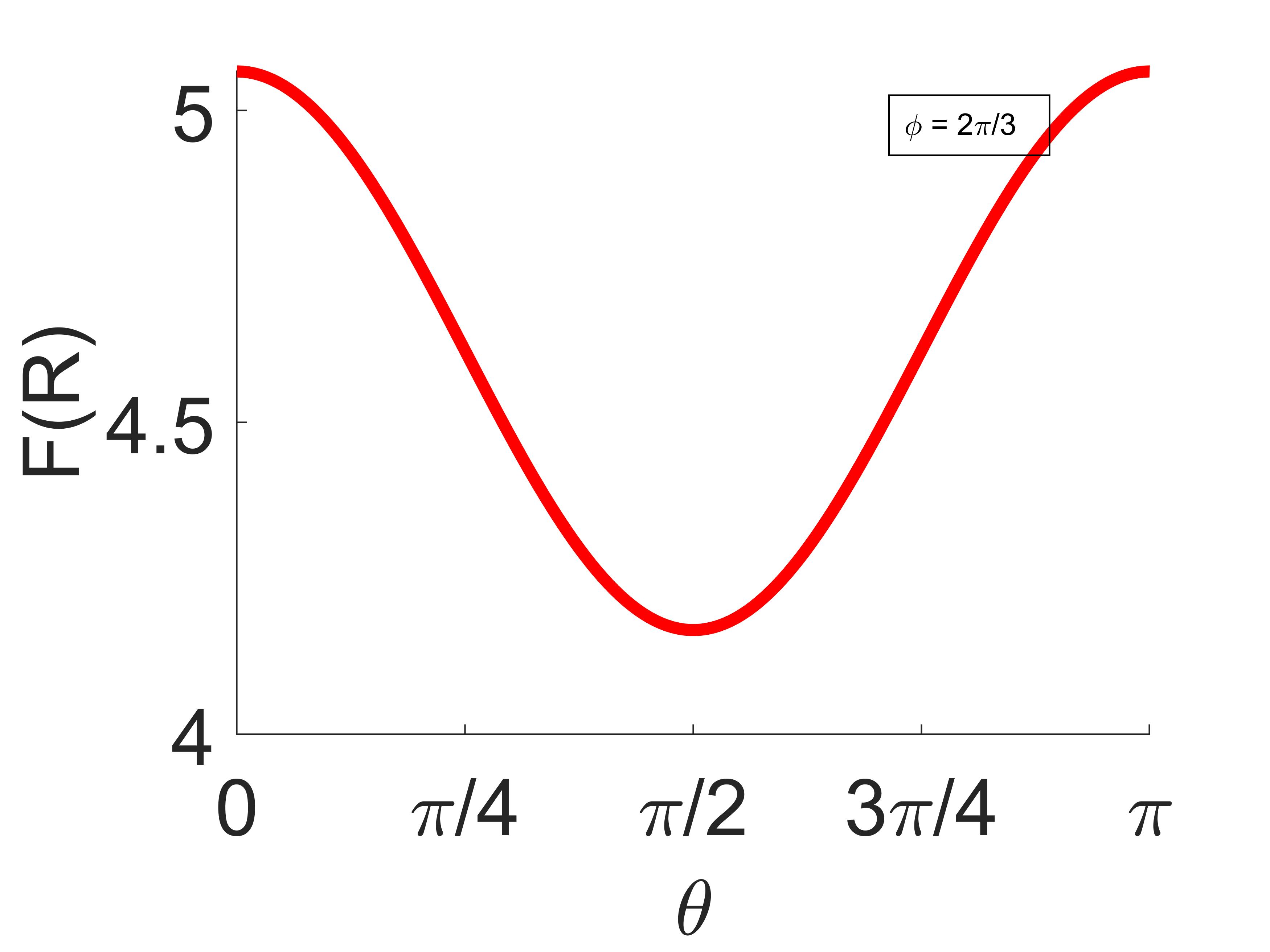}}
\par\end{centering}
\caption{\label{Fig: trsanalytical-1-1}IR breaking WSM: Plot of Range function
w.r.t. (a) $R_{z}$ when the tilt $\mathbf{w}\parallel\hat{z},$; (b) the angle
$\theta=\angle(\mathbf{w},\mathbf{R})$ when $\mathbf{w}\parallel\hat{z}$.}
\end{figure}

\textit{
\begin{align}
\zeta_{IR}^{(2)}\left(\eta,-\eta,R\right) & =-\text{Re}\int_{k=0}^{k_{F\eta}}d\mathbf{k}\text{ }\int_{k'=0}^{\infty}d\mathbf{k}'\frac{\left(\mathbf{w}\cdot\left(\mathbf{k}'+\mathbf{k}\right)\right)^{2}e^{i\left(\mathbf{k}'-\mathbf{k}\right)\cdot\mathbf{R}}}{\left(\hbar v_{F}\left(k'-k\right)\mp2Q_{0}\right)^{3}}
\end{align}
}using, $\mathbf{R}\cdot k=Rx$, $\mathbf{R}\cdot k'=Rx'$
and $\int d\mathbf{k}=\int\text{d}k\text{ }k^{2}\int\text{d\ensuremath{\theta}}\sin\theta\int\text{d}\phi=\int\text{d}k\text{ }k^{2}\int_{-1}^{1}\text{d}x\int\text{d}\phi$,

\begin{align}
\zeta_{IR}^{(2)}\left(\eta,-\eta,R\right) & =-\text{Re}\left[\int_{k=0}^{k_{F\eta}}\text{d}k\text{ }k^{2}\int_{-1}^{1}\text{d}x\int\text{d}\phi\right]\text{ }\left[\int_{k'=0}^{\infty}\text{d}k\text{' }k'^{2}\int_{-1}^{1}\text{d}x'\int\text{d}\phi'\right]\frac{\left(\mathbf{w}\cdot\left(\mathbf{k}'+\mathbf{k}\right)\right)^{2}e^{i\left(\mathbf{k}'-\mathbf{k}\right)\cdot\mathbf{R}}}{\left(\hbar v_{F}\left(k'-k\right)\mp2Q_{0}\right)^{3}},
\end{align}
Now, rescale using, $\mathbf{k}\rightarrow\left(\mathbf{k}-\eta q_{0}\hat{k}\right)$
and $\mathbf{k}'\rightarrow\left(\mathbf{k}'+\eta q_{0}\hat{k'}\right)$
such that $\text{ }\int_{0}^{k_{F\pm}}dk\rightarrow\int_{0}^{k_{F}}dk$
implies $k_{F}=k_{F\pm}\pm q_{0}$. As a result we rewrite,

\textit{
\begin{align}
 & \zeta_{IR}^{(2)}\left(\eta,-\eta,R\right)\nonumber \\
 & =-\text{Re}\left[\int_{k=0}^{k_{F}}\text{d}k\text{ }\left(k-\eta q_{0}\right)^{2}\int_{-1}^{1}\text{d}x\int\text{d}\phi\right]\text{ }\left[\int_{k'=k_{F}}^{\infty}\text{d}k\text{' }\text{ }\left(k+\eta q_{0}\right)^{2}\int_{-1}^{1}\text{d}x'\int\text{d}\phi'\right]\nonumber \\
 & \frac{\left[\left(\mathbf{w}\cdot\left(\mathbf{k}'+\eta q_{0}\hat{k'}\right)\right)^{2}+\left(\mathbf{w}\cdot\left(\mathbf{k}-\eta q_{0}\hat{k}\right)\right)^{2}+2\left(\mathbf{w}\cdot\left(\mathbf{k}'+\eta q_{0}\hat{k'}\right)\right)\left(\mathbf{w}\cdot\left(\mathbf{k}-\eta q_{0}\hat{k}\right)\right)\right]}{\left(k'-k\right)^{3}}\nonumber \\
 & \times e^{i\left(\mathbf{k}'+\eta q_{0}\hat{k'}-\mathbf{k}+\eta q_{0}\hat{k}\right)\cdot\mathbf{R}}.
\end{align}
}
Now, using $\mathbf{w}=w_{x}\hat{x}+w_{y}\hat{y}+w_{z}\hat{z}$ and
$\mathbf{R}=\hat{z}R$,
\textit{
\begin{align}
\text{ } & \zeta_{IR}^{(2)}\left(\eta,-\eta,R\right)\nonumber \\
 & =\text{Re}\left(\left(\frac{\mathbf{w}}{v_{F}}\cdot\frac{\partial}{\partial\mathbf{R}}\right)^{2}\left[\int_{k=0}^{k_{F}}\text{d}k\text{ }\left(k-\eta q_{0}\right)^{2}\int_{-1}^{1}\text{d}x\int\text{d}\phi\right]\text{ }\text{ }e^{-i\left(\mathbf{k}-\eta q_{0}\hat{k}\right)\cdot\mathbf{R}}\right)\nonumber \\
 & \left[\int_{k'=k_{F}}^{\infty}\text{d}k\text{' }\text{ }\left(k+\eta q_{0}\right)^{2}\int_{-1}^{1}\text{d}x'\int\text{d}\phi'\right]\frac{e^{i\left(\mathbf{k}'+\eta q_{0}\hat{k'}\right)\cdot\mathbf{R}}}{\left(k'-k\right)^{3}}\nonumber \\
 & +\text{Re}\left[\int_{k=0}^{k_{F}}\text{d}k\text{ }\left(k-\eta q_{0}\right)^{2}\int_{-1}^{1}\text{d}x\int\text{d}\phi\text{ }\text{ }e^{-i\left(\mathbf{k}-\eta q_{0}\hat{k}\right)\cdot\mathbf{R}}\right]\nonumber \\
 & \left[\left(\frac{\mathbf{w}}{v_{F}}\cdot\frac{\partial}{\partial\mathbf{R}}\right)^{2}\left[\int_{k'=k_{F}}^{\infty}\text{d}k\text{' }\text{ }\left(k+\eta q_{0}\right)^{2}\int_{-1}^{1}\text{d}x'\int\text{d}\phi'\right]\frac{e^{i\left(\mathbf{k}'+\eta q_{0}\hat{k'}\right)\cdot\mathbf{R}}}{\left(k'-k\right)^{3}}\right]\nonumber \\
 & +\text{Re}\left[\left(\frac{\mathbf{w}}{v_{F}}\cdot\frac{\partial}{\partial\mathbf{R}}\right)\left[\int_{k=0}^{k_{F}}\text{d}k\text{ }\left(k-\eta q_{0}\right)^{2}\int_{-1}^{1}\text{d}x\int\text{d}\phi\text{ }\text{ }e^{-i\left(\mathbf{k}-\eta q_{0}\hat{k}\right)\cdot\mathbf{R}}\right]\right]\nonumber \\
 & \left[\left(\frac{\mathbf{w}}{v_{F}}\cdot\frac{\partial}{\partial\mathbf{R}}\right)\left[\int_{k'=k_{F}}^{\infty}\text{d}k\text{' }\text{ }\left(k+\eta q_{0}\right)^{2}\int_{-1}^{1}\text{d}x'\int\text{d}\phi'\right]\frac{e^{i\left(\mathbf{k}'+\eta q_{0}\hat{k'}\right)\cdot\mathbf{R}}}{\left(k'-k\right)^{3}}\right].
\end{align}
}
Writting, $\mathbf{R}\cdot k=Rx$ and $\mathbf{R}\cdot k'=Rx'$
and integrating over $x,x'$ we obtain,
\textit{
\begin{align}
 & \zeta_{IR}^{(2)}\left(\eta,-\eta,R\right)\nonumber \\
 & =\text{Re}\left(\int_{k=0}^{k_{F}}\text{d}k\text{ }\left(\frac{\mathbf{w}}{v_{F}}\cdot\frac{\partial}{\partial\mathbf{R}}\right)^{2}\frac{4\pi(k-\eta\text{\ensuremath{q_{0}}})\sin(R(k-\eta\text{\ensuremath{q_{0}}}))}{R}\right)\left[\int_{k'=k_{F}}^{\infty}\text{d}k\text{' }\frac{4\pi(k'+\eta\text{\ensuremath{q_{0}}})\sin(R(k'+\eta\text{\ensuremath{q_{0}}}))}{R\left(k'-k\right)^{3}}\right]\nonumber \\
 & +\text{Re}\left[\int_{k=0}^{k_{F}}\text{d}k\text{ }\frac{4\pi(k-\eta\text{\ensuremath{q_{0}}})\sin(R(k-\eta\text{\ensuremath{q_{0}}}))}{R}\right]\left(\frac{\mathbf{w}}{v_{F}}\cdot\frac{\partial}{\partial\mathbf{R}}\right)^{2}\int_{k'=k_{F}}^{\infty}\text{d}k\text{' }\text{ }\frac{4\pi(k'+\eta\text{\ensuremath{q_{0}}})\sin(R(k'+\eta\text{\ensuremath{q_{0}}}))}{R\left(k'-k\right)^{3}}\nonumber \\
 & +\text{Re}\left[\int_{k=0}^{k_{F}}\text{d}k\text{ }\left(\frac{\mathbf{w}}{v_{F}}\cdot\frac{\partial}{\partial\mathbf{R}}\right)\frac{4\pi(k-\eta\text{\ensuremath{q_{0}}})\sin(R(k-\eta\text{\ensuremath{q_{0}}}))}{R}\right]\left(\frac{\mathbf{w}}{v_{F}}\cdot\frac{\partial}{\partial\mathbf{R}}\right)\left[\int_{k'=k_{F}}^{\infty}\text{d}k\text{' }\text{ }\frac{4\pi(k'+\eta\text{\ensuremath{q_{0}}})\sin(R(k'+\eta\text{\ensuremath{q_{0}}}))}{R\left(k'-k\right)^{3}}\right].
\end{align}
}

\section{Finite tilt limit:}
\begin{figure}
\begin{centering}
\includegraphics[width=0.5\linewidth]{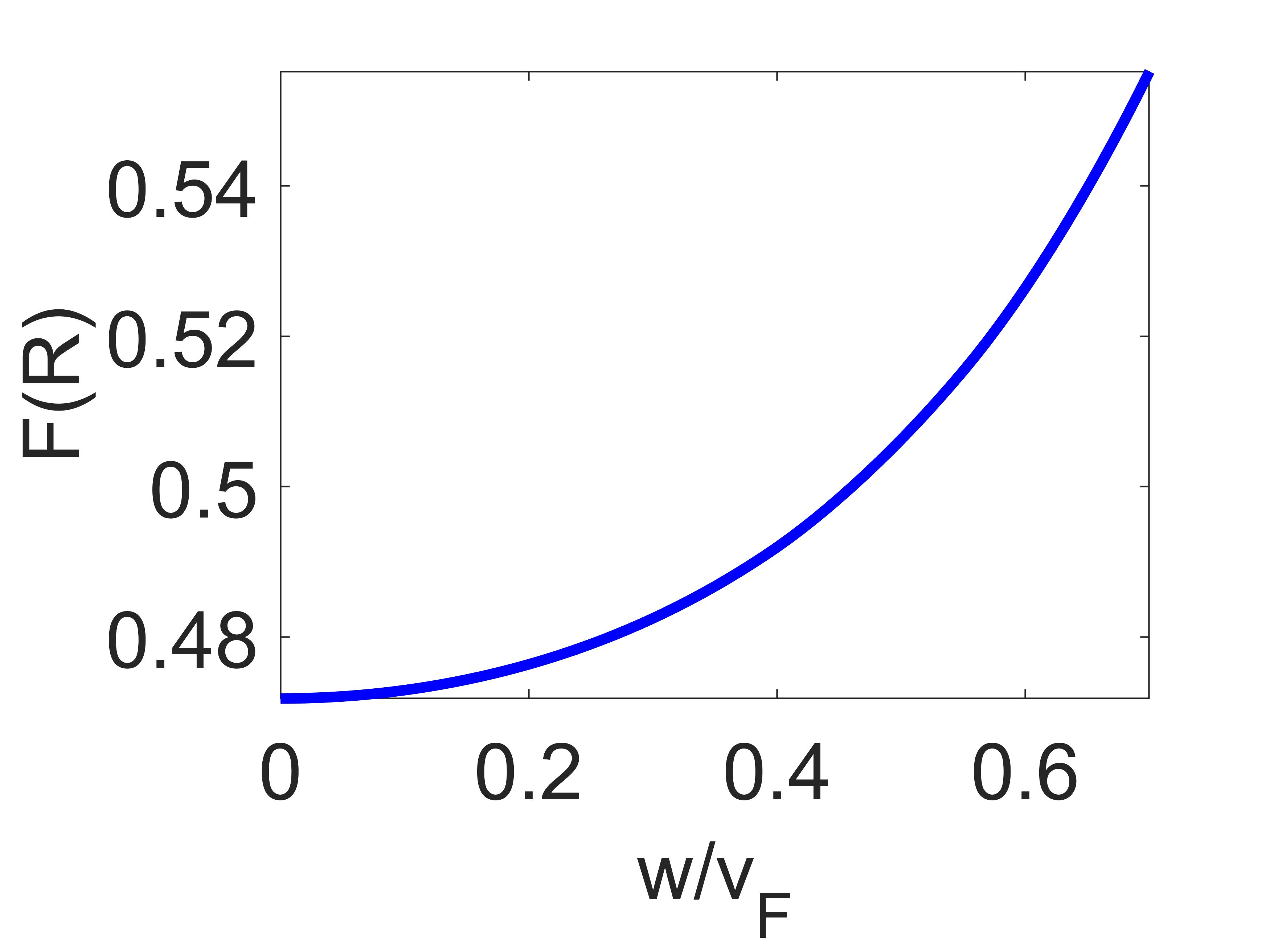}
\par\end{centering}
\caption{\label{Fig: trsnumerical} Plot of Range function w.r.t. tilt magnitude
$\mathbf{w}$ for TRS breaking WSMs}

\end{figure}
Thus far, we have considered the analytical expression of the RKKY
range function in the small-tilt limit. In this section, we will analyze
the RKKY range function numerically for any arbitrary tilt value.
We will consider the general expressions in Eqs. (\ref{eq:arb tilt 1}-\ref{eq:arb tilt 3})
and assume parallel tilt vector for all Weyl nodes. In our calculations,
we arbitrarily set the tilt as $\mathbf{w}\parallel\hat{z}$ and express
the separation vector $\mathbf{R}=\mathrm{R}(\sin(\theta)\cos(\phi)\hat{x}+\sin(\theta)\sin(\phi)\hat{y}+\cos(\theta)\hat{z})$.
In Fig. \ref{Fig: trsnumerical}, we plot the range function with
respect to the tilt magnitude $|\mathbf{w}|$ for TRS broken case.
The RKKY function has non-linear dependence on tilt, matching the
trend of our analytical result.